\documentclass{elsart}

\usepackage{natbib}

\usepackage{graphicx}
\usepackage{amssymb}
\usepackage{amsmath}

\newcommand{\bi}[1]{\boldsymbol{#1}}
\newcommand{\mc}[1]{\mathcal{#1}}
\newcommand{\mf}[1]{\mathfrak{#1}}
\newcommand{\bb}[1]{\mathbb{#1}}

\renewcommand{\atop}[2]{\genfrac{}{}{0pt}{}{#1}{#2}}
\newcommand{\order}[0]{\mathcal{O}\left(\frac{1}{N}\right)}

\newcounter{thmno}
\begin{document}

\begin{frontmatter}



\title{Error thresholds in a mutation--selection model with
  Hopfield-type fitness}


\author{Tini Garske}

\address{Fakult\"at f\"ur Mathematik, Universit\"at Bielefeld,
  Postfach 100131, D-33501 Bielefeld, Germany\\
  and\\
  Applied Maths Department, Faculty of Mathematics and Computing, The 
  Open University, Walton Hall, Milton Keynes, MK7 6AA, UK}
\ead{t.garske@open.ac.uk, tini@funandgames.org}

\begin{abstract}
  A deterministic mutation--selection model in the sequence space
  approach is investigated. Genotypes are identified with two-letter
  sequences. Mutation is modelled as a Markov process, fitness
  functions are of Hopfield type, where the fitness of a sequence is
  determined by the Hamming distances to a number of predefined
  patterns. Using a maximum principle for the population mean fitness
  in equilibrium, the error threshold phenomenon is studied for
  quadratic Hopfield-type fitness functions with small numbers of 
  patterns. Different from previous investigations of the Hopfield
  model, the system shows error threshold behaviour not for all
  fitness functions, but only for certain parameter values.
\end{abstract}


\end{frontmatter}

\section{Introduction}
\label{Introduction}

Population genetics is concerned with the investigation of the genetic
structure of populations, which is influenced by evolutionary factors
such as mutation, selection, recombination, migration and genetic
drift. For excellent reviews of the theoretical aspects of this
field, see \citet{BG00,CK70,Ewe04}.

In this paper, the antagonistic interplay of mutation and selection
shall be investigated, with mutation generating the genetic variation
upon which selection can act. Pure mutation--selection models exclude
genetic drift and are therefore deterministic models, and accurate
only in the limit of an infinite population size \citep[for a review,
see][]{Bur00}. A further simplification taken here is to consider only
{\em haploid} populations, where the genetic material exists in one
copy only in each cell. However, the equations used here to describe
evolution apply as well to diploid populations without dominance.

For the modelling of the types considered, the {\em sequence space
  approach} is used, which has first been used by \cite{Eig71} to
model the structure of the DNA, where individuals are taken to be
sequences. Here, the sequences shall be written in a two-letter
alphabet, thus simplifying the full four-letter structure of DNA
sequences. In this approach, the modelling is based on the microscopic
level, at which the mutations occur, hence the mutation process is
fairly straightforward to model. However, the modelling of selection
is a more challenging task, as selection acts on the phenotype, and
the mapping from genotype to phenotype is by no means simple. To this
end, the concept of the {\em fitness landscape} \citep{KL87} is
introduced as a function on the sequence space, assigning to each
possible genotype a fitness value which determines the reproduction
rate. Apart from the problem that a realistic fitness landscape would
have to be highly complex (too complex for a mathematical treatment),
there is also very limited information available concerning the nature
of realistic fitness functions.  Therefore, the modelling of fitness
is bound by feasibility, trying to mimic general features that are
thought to be essential for realistic fitness landscapes such as the
degree of ruggedness.

A very common type of fitness functions is the class of
permutation-invariant fitness functions, where the fitness of a
sequence is determined by the number of mutations it carries compared
to the wild-type, but not on the locations of the mutations within the
sequence. Although this model describes the accumulation of small
mutational effects surprisingly well, it is a simplistic model that
lacks a certain degree of ruggedness that is thought to be an
important feature of realistic fitness landscapes \citep{BG00}.

In this paper, Hopfield-type fitness functions \citep{Hop82} are
treated as a more complex model. Here, the fitness of a sequence is
not only determined by the number of mutations compared to one
reference sequence, but to a number of predefined sequences, the {\em
patterns}. This yields a class of fitness landscapes that contain a
higher degree of ruggedness, which can be tuned by the number of
patterns chosen. While this can still be treated with the methods used
here, it is a definite improvement on the restriction of
permutation-invariant fitness functions.

Particular interest is taken in the phenomenon of mutation driven
error thresholds, where the population in equilibrium changes from
viable to non-viable within a narrow regime of mutation rates. In this
paper, a few examples of Hopfield-type fitness functions are
investigated with respect to the error threshold phenomenon.

Section \ref{The mutation--selection model in sequence space}
introduces the basic mutation--selection model with its main
observables. In section \ref{Sequences as types}, the model is applied
to the sequence space approach, formulating the mutation and fitness
models explicitly. Sections \ref{Lumping for the Hopfield-type
  fitness} and \ref{The maximum principle} present the method, which
relies on a lumping of the large number of sequences into classes on
which a coarser mutation--selection process is formulated. This
lumping is necessary to formulate a simple maximum principle to
determine the population mean fitness in equilibrium. In section
\ref{Error thresholds}, this maximum principle is used to investigate
some examples of Hopfield-type fitness functions with respect to the
error threshold phenomenon.

\section{The mutation--selection model}
\label{The mutation--selection model in sequence space}

The model used here (as detailed below) is a pure mutation--selection
model in a time-continuous formulation as used by \citet{HRWB02} and
\citet{GG04a,GG04b}, for instance.

\paragraph*{Population.}
The evolution of a population where the only evolutionary forces are
mutation and selection is considered, thus excluding other factors
such as drift or recombination for instance. Individuals in the
population shall be assigned a type $i$ from the finite {\em type
  space} $\mf{S}$. The population at any time $t$ is described by the
{\em population distribution} $\bi{p}(t)$, a vector of dimension
$|\mf{S}|$, the cardinality of the type space. An entry $p_i(t)$ gives
the fraction of individuals in the population that are of type $i$.
Thus the population is normalised such that $\sum_i p_i(t)=1$.

\paragraph*{Evolutionary processes.}
The evolutionary processes that occur are birth, death and mutation
events. Birth and death events occur with rates $b_i$ and $d_i$ that
depend on the type $i$ of the individual in question, and taken
together, they give the effective reproductive rate, or {\em fitness}
as $r_i=b_i-d_i$. Mutation from type $i$ to type $j$ depends on both
initial and final type and happens with rate $m_{ji}$.  These rates
are conveniently collected in square matrices $\mc{R}$ and $\mc{M}$ of
dimension $|\mf{S}|$, where the reproduction or fitness matrix
$\mc{R}$ with entries $r_i$ is diagonal. The off-diagonal entries of
the mutation matrix $\mc{M}$ are given by the mutation rates $m_{ji}$,
and as mutation does not change the number of individuals, the
diagonal entries of $\mc{M}$ are chosen such that
$\mc{M}_{ii}=-\sum_{j\neq i}\mc{M}_{ji}$, which makes $\mc{M}$ a
Markov generator. The time evolution operator $\mc{H}$ is given by the
sum of reproduction and mutation matrix, $\mc{H}=\mc{R}+\mc{M}$.

\paragraph*{Deterministic evolution equation.}
In the deterministic limit of an infinite population size, the
evolution of the population is governed by the evolution equation
\begin{equation}
\label{evolution equation}
\bi{\dot{p}}(t)=\left[\mc{H}-\bar{r}(t)\bi{1}\right]\bi{p}(t) \;,
\end{equation}
where $\bar{r}(t)=\sum_i r_i p_i(t)$ is the population mean fitness.
The term with $\bar{r}$ is needed to preserve the normalisation of the
population. Note that this term makes the evolution equation
(\ref{evolution equation}) nonlinear.

\paragraph*{Equilibrium.}
The main interest focuses on the equilibrium, i.e., the behaviour if
$\bi{\dot{p}}=0$, which is attained for $t\rightarrow\infty$. All
equilibrium quantities shall be denoted by omitting the argument $t$,
for instance the equilibrium population distribution is $\bi{p}$. In
equilibrium, the evolution equation (\ref{evolution equation}) becomes
an eigenvalue equation for $\mc{H}$, with leading eigenvalue $\bar{r}$
and corresponding eigenvector $\bi{p}$. If $\mc{M}$ is irreducible, as
shall be assumed throughout, Perron-Frobenius theory \citep[see, for
instance,][appendix]{Kar66} applies, which guarantees that the leading
eigenvalue $\bar{r}$ of $\mc{H}$ is non-degenerate and the
corresponding right eigenvector $\bi{p}$ is strictly positive, which
implies that it can be normalised as a probability distribution.

\paragraph*{Ancestral distribution.}
Similarly to the population distribution $\bi{p}$, there is also
another important distribution in this model, namely the ancestral
distribution $\bi{a}(\tau,t)$. Consider the population at time
$t+\tau$, but count each individual not as its current type, but as
the type its ancestor had at time $t$. Thus an entry of the ancestral
distribution $a_i(\tau,t)$ determines the fraction of the population
at $t+\tau$ whose ancestor at time $t$ was of type $i$. In the limit
$\tau,t\rightarrow\infty$, this also approaches an equilibrium
distribution $\bi{a}$. As shown by \citet{HRWB02}, the equilibrium
ancestral distribution can be obtained as a product of the left and
right PF (Perron-Frobenius) eigenvectors $\bi{z}$ and $\bi{p}$ of the
time-evolution operator $\mc{H}$ as $a_i=z_i p_i$, where $\bi{z}$ is
normalised such that $\sum_i a_i=1$.

\paragraph*{Population and ancestral means.}
Any function on the type space given, say, by $f=(f_i)_{i\in\mf{S}}$
can be averaged with respect to the population or the ancestral
distribution distribution. The population mean of $f$ is given by
\begin{equation}
\label{population mean}
\bar{f}(t)=\sum_i f_i p_i(t) \;,
\end{equation}
whereas the ancestral mean is
\begin{equation}
\label{ancestral mean}
\hat{f}(\tau,t)=\sum_i f_i a_i(\tau,t) \;.
\end{equation}
Note that the time-dependence of the means only comes from the
distribution, whereas the function $f$ is considered constant in time.
In equilibrium, time dependence is again omitted such that the
equilibrium population and ancestral means are denoted $\bar{f}$ and
$\hat{f}$, respectively.  An important example of the population mean
is the population mean fitness $\bar{r}(t)$ from equation
(\ref{evolution equation}).

\section{Sequences as types}
\label{Sequences as types}

In the previous section, the types are a rather abstract concept. In
order to formulate the particular mutation and fitness models, they
shall now be specified as sequences, mimicking the structure of the
DNA \citep[cf.][]{Eig71}. For simplicity, only two-state sequences are
considered, i.e., sequences that have at each site one out of two
possible entries. However, the method used here can immediately be
generalised to a more realistic four-state model \citep[see][]{Gar05}.

The types therefore are associated with sequences $\bi{\sigma} =
\sigma_1 \sigma_2 \ldots \sigma_N$ of fixed length $N$, written in the
{\em alphabet} $\mc{A}=\{0,1\}$, thus $\sigma_{\alpha} \in \mc{A}$ for
$\alpha = 1, \ldots, N$. This means that there are $2^N$ different
sequences, and thus the type space (or {\em sequence space}) $\mf{S}$
has cardinality $|\mf{S}|=2^N$.

\subsection{Mutation model} 

A simple mutation model that neglects any processes changing the
length of the sequence, such as deletions or insertions, is used.
Mutations are modelled as point processes, where an arbitrary site
$\sigma_{\alpha}$ is switched with rate $\mu$, such that the mutation
rate between sequences that differ only in one particular site is
given by $\mu/N$.  Sequences that differ in more than one site cannot
mutate into one another within a single mutational step. This is known
as the {\em single step mutation model}, introduced by \citet{OK73}.

The mutation model defines a neighbourhood in the sequence space
\citep{RS02}.  A convenient measure for the distance between sequences
is the Hamming distance $d_H(\bi{\sigma}, \bi{\sigma'})$, which counts
the number of sites at which the sequences $\bi{\sigma}$ and
$\bi{\sigma'}$ differ \citep{Ham50,vLi82}. With this, the mutation
matrix is explicitly given as
\begin{equation}
\label{mutation matrix}
\mc{M}_{\bi{\sigma \sigma'}}=
\begin{cases}
\mu/N & \text{if $d_H(\bi{\sigma}, \bi{\sigma'})=1$\,,}\\
0     & \text{if $d_H(\bi{\sigma}, \bi{\sigma'})>1$\,,}\\
-\mu  & \text{if $\bi{\sigma} = \bi{\sigma'}$\,.}
\end{cases}
\end{equation}
The diagonal entry is chosen such that $\mc{M}$ fulfils the Markov
condition $\sum_j \mc{M}_{ji} = 0$.

\subsection{Fitness functions}
A rather simple, though commonly used type of fitness function is the
{\em per\-mu\-ta\-tion-in\-vari\-ant fitness}. There, the fitness of a
sequence depends only on the number of mutations it has compared to a
reference type, not on their position along the sequence.  Thus
fitness is a function of the Hamming distance to the reference
sequence, which is usually chosen as the wild-type.  The Hamming
distance to the wild-type is also called the {\em mutational distance}
$d$. 

A non-permutation-invariant fitness that contains some ruggedness, but
is simple enough to be dealt with in this framework, is the {\em
  Hopfield-type fitness}, a special type of spin-glass model, which
has been introduced by \citet{Hop82} as a model for neural networks.
Instead of comparing a sequence only to the wild-type, as it is done
for the permutation-invariant fitness, the Hopfield-type fitness of a
sequence is determined by its Hamming distances to $p+1$ reference
sequences, the {\em patterns} $\bi{\xi}^q$, $q=0, \ldots, p$. The
Hopfield-type fitness shall be defined in terms of the {\em specific
  distances} $w^q$, which are the Hamming distances with respect to
the patterns,
\begin{equation}
\label{def specific distances}
w^q=d_H(\bi{\xi}^q, \bi{\sigma})\,,
\end{equation}
and thus the fitness is given as  
\begin{equation}
\label{Hopfield-type fitness}
r_{\bi{\sigma}} = 
r_{\bi{\sigma}}((w^q)_{q=0,\ldots,p})\,.
\end{equation}
Note that in the case of a single pattern ($p=0$), this yields again a
permutation-invariant fitness.

\section{Lumping for the Hopfield-type fitness}
\label{Lumping for the Hopfield-type fitness}

\subsection{The general lumping procedure}

One problem of the sequence space approach is the large number of
types, which grows exponentially with the sequence length $N$,
$|\mf{S}|=2^N$. The time-evolution operator $\mc{H}$ is a matrix of
size $|\mf{S}|\times|\mf{S}|$, and in this set-up one is interested in
its leading eigenvalue $\bar{r}$ and the corresponding right and left
eigenvectors $\bi{p}$ and $\bi{z}$.

The relevant sequence length depends on the particular application one
has in mind, but it is typically rather long. If one aims to model the
whole genome of a virus or a bacterium, $N$ has to be in the region of
$N\approx 10^6$, but even a single gene has of the order of $N\approx
10^3$ base pairs. These values lead to matrices of a size that makes
the eigenvalues and eigenvectors inaccessible.

For some types of fitness functions, this problem can be reduced by
{\em lumping} together types into {\em classes} of types, and
considering the new process on a reduced sequence space, which
contains the classes rather than the individual types. Under certain
circumstances, mutation is described as a Markov process in the
emerging lumped process as well, such that this process is accessible
to Markov process methods, and the framework developed in section
\ref{The mutation--selection model in sequence space} can directly be
applied to the lumped system.

The lumping of the mutation process is a standard procedure in the
theory of Markov chains \citep[chapter 6]{KS60}, see also
\citet{BBBK05} for an application to mutation--selection models. This
lumping leads to a meaningful mutation--selection model on the reduced
type space, if all sequences lumped together into one class have the
same fitness.

\begin{figure}
\setlength{\unitlength}{1pt}
\begin{center}
\begin{picture}(258,171)
\put(15, 5){\makebox(0,0){$\mf{S}_k$}}
\put(200, 30){\makebox(0,0){$\mf{S}_{\ell}$}}
\put(23, 90){\makebox(0,0){$i_1$}}
\put(25, 40){\makebox(0,0){$i_2$}}
\includegraphics[width=0.6\textwidth]{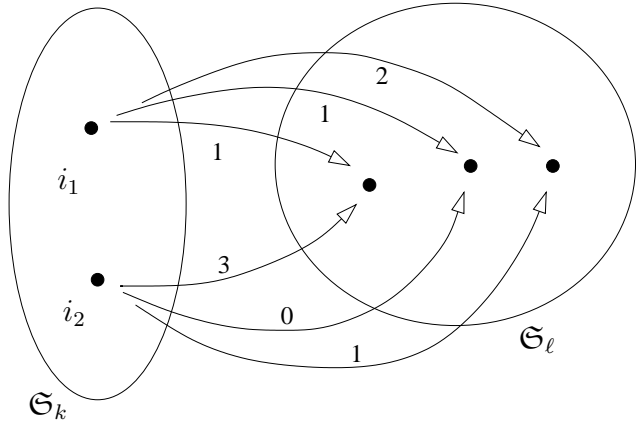}
\end{picture}
\caption[Visualisation of the compatibility with lumping.]
{\label{Fig lumping}Visualisation of the compatibility with lumping:
  Consider two classes $\mf{S}_k$ and $\mf{S}_{\ell}$. The mutation
  rates from the types in $\mf{S}_k$ to the types in $\mf{S}_{\ell}$
  (given next to the arrows) are compatible with a lumping with
  respect to $\mf{S}_k$ and $\mf{S}_{\ell}$, because the sum of the
  mutation rates from type $i_1$ to all types in $\mf{S}_{\ell}$ is
  given by $u_{\mf{S}_{\ell},{i_1}}=1+1+2=4$, which is identical with
  those from $i_2$, $u_{\mf{S}_{\ell},{i_2}}=3+0+1=4$. }
\end{center}
\end{figure}

It is possible to lump the Markov chain given by the mutation matrix
$\mc{M}$ with state space $\mf{S}$ with respect to a particular
partition $\mf{S} =\dot{\bigcup}_{k=0}^r \mf{S}_k$, if and only if for
each pair $\mf{S}_k,\mf{S}_{\ell}$ the cumulative mutation rates
\begin{equation}
\label{general cumulative mutation rates}
u_{\mf{S}_{\ell},i}:=\sum_{j\in\mf{S}_{\ell}}\mc{M}_{ji}
\end{equation}
 from type $i\in\mf{S}_k$ into $\mf{S}_{\ell}$, are identical for all
$i\in\mf{S}_k$, cf.\ the example shown in figure \ref{Fig lumping}.
In this case, the lumped process, with $r+1$ states
$\mf{S}_0,\ldots,\mf{S}_r$ and mutation rates $u_{{\mf{S}_{\ell}},
{i}}$ for any $i\in\mf{S}_k$, is again a Markov chain \citep[theorem
6.3.2]{KS60}.

\subsection{Defining the partial distances}

Whereas in the case of a permutation-invariant fitness function, the
lumping procedure is fairly simple, collecting all sequences with the
same Hamming distance to the wild-type into classes, and considering
cumulative mutation rates between these classes, the lumping for the
Hopfield-type fitness is somewhat more complex. For a two-state model
with Hopfield fitness, it has been performed for instance by
\citet{BBBK05}, and this shall be recollected in the remainder of this
section.

First, the quantities with respect to which the lumping shall be
performed must be defined. To this end, consider as an example the
case of sequence length $N=12$ with three patterns ($p=2$), and let
the patterns $\bi{\xi}^q$ be collected in a $(p+1)\times N$ matrix
$\bi{\xi}$, such that the $q$th row of $\bi{\xi}$ is pattern
$\bi{\xi}^q$. Without loss of generality, the pattern $\bi{\xi}^0$ can
always be chosen as $\bi{\xi}^0=000\ldots0$.  Let the patterns in this
example be given as
\begin{equation}
\label{example pattern}
\setcounter{MaxMatrixCols}{12}
\bi{\xi} =
\begin{pmatrix} \bi{\xi}^0 \\ \bi{\xi}^1 \\ \bi{\xi}^2 \end{pmatrix}
= \begin{pmatrix}
0 & 0 & 0 & 0 & 0 & 0 & 0 & 0 & 0 & 0 & 0 & 0 \\
1 & 0 & 1 & 1 & 0 & 1 & 0 & 1 & 1 & 1 & 0 & 0 \\
1 & 0 & 0 & 0 & 1 & 1 & 1 & 1 & 1 & 1 & 1 & 0 
\end{pmatrix} \,.
\end{equation}
Note that there are only $2^p$ different types of sites (corresponding
to the columns of $\bi{\xi}$). These are collected in a $(p+1) \times
2^p$ matrix $\bi{\rho}$, the columns of which correspond to the
possible types of sites in the matrix of patterns $\bi{\xi}$. For the
case $p=2$, this $3\times4$ matrix is given as
\begin{equation}
\label{rho}
\bi{\rho}=
\begin{pmatrix}
0&0&0&0\\
0&0&1&1\\
0&1&0&1
\end{pmatrix} \,.
\end{equation}
Using the column vectors $\bi{\rho}_v$ of the matrix $\bi{\rho}$, the
patterns given in equation (\ref{example pattern}) can alternatively
be expressed as
\begin{equation}
\label{example pattern with rho}
\bi{\xi}=\left(
\bi{\rho}_4, \bi{\rho}_1, \bi{\rho}_3, \bi{\rho}_3, 
\bi{\rho}_2, \bi{\rho}_4, \bi{\rho}_2, \bi{\rho}_4, 
\bi{\rho}_4, \bi{\rho}_4, \bi{\rho}_2, \bi{\rho}_1\right)\,,
\end{equation}
classifying the sites into $2^p$ classes according to which of the
column vectors $\bi{\rho}_v$ of $\bi{\rho}$ coincides with the column
vector $\bi{\xi}_{\alpha}$ of the patterns $\bi{\xi}$ at site
$\alpha$.  

Let $\Lambda=\{1,\ldots,N\}$ be the index set of sites, with a
partition into $2^p$ subsets $\Lambda_v$ induced by the patterns
$\bi{\xi}$ such that
\begin{equation}
\Lambda=\dot{\bigcup}_{v=1}^{2^p}\Lambda_v\,,
\end{equation}
with
\begin{equation}
\alpha\in\Lambda_v \quad\Longleftrightarrow\quad
\bi{\xi}_{\alpha}=\bi{\rho}_v \,.
\end{equation}
Patterns can be characterised by the number of sites $N_v=|\Lambda_v|$
in each subset.  The example patterns (\ref{example pattern}) can
therefore be described by
\begin{align}
\label{example pattern partition}
\Lambda_1&=\{2,12\}\,, & \Lambda_2&=\{5,7,11\}\,, 
& \Lambda_3&=\{3,4\}\,, & \Lambda_4&=\{1,6,8,9,10\}\,,\notag\\
N_1&=2\,, & N_2&=3\,, & N_3&=2\,, & N_4&=5\,.
\end{align}

Considering only subsequences $\bi{\sigma}_v :=
(\sigma_{\alpha})_{\alpha \in \Lambda_v}$, the {\em partial distances}
$d_v$ of a sequence $\bi{\sigma}$ with respect to the pattern
$\bi{\xi}^0$ are defined as the Hamming distance between $\bi{\sigma}$
and $\bi{\xi}^0$, {\em restricted to the subsequence
$\bi{\sigma}_v$}. Therefore, they can be written as
\begin{equation}
\label{partial distances}
d_v:=\sum_{\alpha\in\Lambda_v} \delta(1,\sigma_{\alpha}) 
\quad\mbox{with}\quad 0 \le d_v \le N_v \,,
\end{equation}
such that the specific distance with respect to pattern $\bi{\xi}^0$
is given by $w^0 = \sum_{v=1}^{2^p} d_v$.

Because the differences between each of the patterns $\bi{\xi}^q$
within the subsets $\Lambda_v$ of $\Lambda$ are known (and recorded in
the matrix $\bi{\rho}$), it is sufficient to consider only the partial
distances $d_v$ with respect to one pattern, here $\bi{\xi}^0$; the
partial distances $d_v^q$ with respect to any other pattern can be
expressed in terms of the $d_v$ as
\begin{equation}
\label{q-partial distances}
d_v^q=\sum_{\alpha\in\Lambda_v} \delta(1-\rho_v^q,\sigma_{\alpha})
=\begin{cases}
d_v     & \text{if $\rho_v^q=0$,}\\
N_v-d_v & \text{if $\rho_v^q=1$,}\\
\end{cases}
\end{equation}
using the matrix elements $\rho_v^q$ of $\bi{\rho}$. 

The specific distance $w^q$ to any pattern $\bi{\xi}^q$ can be
expressed as
\begin{equation}
\label{specific distances}
w^q
=\sum_{v=1}^{2^p}\sum_{\alpha\in\Lambda_v} 
       \delta(1-\rho_v^q, \sigma_{\alpha})
=\sum_{v\in V_0^q}d_v+\sum_{v\in V_1^q}(N_v-d_v) \,,
\end{equation}
where the index set $V=\{1,\ldots,2^p\}$ of classes is partitioned
into two subsets, $V=V_0^q\cup V_1^q$ with $V_0^q=\{v|\rho_v^q=0\}$
and $V_1^q=\{v|\rho_v^q=1\}$.

Hence, by specifying the $2^p$ partial distances $d_v$ with respect to
pattern $\bi{\xi}^0$, the specific distances $w^q$ with respect to any
pattern $\bi{\xi}^q$ are determined, which in turn determine the
fitness. This implies that all sequences with the same partial
distances $d_v$ have the same fitness.  Thus the partial distances
$d_v$ to pattern $\bi{\xi}^0$, collected in a {\em mutational
  distance} vector $\bi{d}=(d_v)_{v=1\ldots2^p}$, shall be the
quantities that label the classes in the lumped system.

\subsection{Lumping with respect to the partial distances}

The relevant partition of the sequence space is given by $\mf{S} =
\dot{\bigcup}_{\bi{d}} \mf{S}_{\bi{d}}$ with $\mf{S}_{\bi{d}} =
\{\bi{\sigma} | \bi{d}_{\bi{\sigma}}=\bi{d} \}$, and the reduced
sequence space, or {\em mutational distance space} $\mc{S}$, contains
the classes $\mc{S}=\{ \bi{d} | 0 \le d_v \le N_v, v = 1 \ldots 2^p
\}$.

Considering again the subsequences $\bi{\sigma}_v $, there are $N_v+1$
possible different $d_v$ (as $d_v$ takes values from $0$ to $N_v$),
and there are $\binom{N_v}{d_v}$ different subsequences
$\bi{\sigma}_v$ for each $d_v$. Hence, considering all sites, there
are
\begin{equation}
\label{number of types}
|\mathcal{S}|=\prod_{v=1}^{2^p}(N_v+1)
\end{equation}
different $\bi{d}$, and
\begin{equation}
\label{nd}
n_{\bi{d}}=\prod_{v=1}^{2^p} \binom{N_v}{d_v}
\end{equation}
sequences $\bi{\sigma}$ that are mapped onto each $\bi{d}$. For the
patterns $\bi{\xi}$ chosen as example (\ref{example pattern}), we have
$|\mc{S}|=3 \cdot 4 \cdot 3 \cdot 6 = 216$, while the full sequence
space has dimension $|\mf{S}| = 2^{12} = 4096$.

In the single step mutation model, the only neighbours of a sequence
$\bi{\sigma}$ with distance vector $\bi{d}$ lie in the classes
$\bi{d}\pm\bi{e}_v$, where the $\bi{e}_v=(\delta_{vw})_{w=1\ldots2^p}$
are the unit vectors of mutation. Thus the only non-zero cumulative
mutation rates are
\begin{equation}
u^{\pm v}_{\bi{d}} 
:= u_{\bi{d} \pm \bi{e}_v, \bi{\sigma}} 
= \sum_{\bi{\sigma}' \in \mf{S}_{\bi{d} \pm \bi{e}_v} }
   \mc{M}_{\bi{\sigma}'\bi{\sigma}} \,.
\end{equation}
As a sequence with $\bi{d}$ has $(N_v-d_v)$ $0$-sites and $d_v$
$1$-sites in $\Lambda_v$, and the single-site mutation rates are
$\mu/N$ for all sites, the cumulative mutation rates are given by
\begin{align}
\label{cumulative mutation rates}
u^{+v}_{\bi{d}}&=\mu \,(N_v-d_v) /N
&&\text{for $\bi{d}\rightarrow\bi{d}+\bi{e}_v$ and}\notag\\
u^{-v}_{\bi{d}}&=\mu \,d_v /N
&&\text{for $\bi{d}\rightarrow\bi{d}-\bi{e}_v$,}
\end{align}
irrespective of the particular order within the subsequences
$\bi{\sigma}_v$. Therefore the cumulative mutation rates $u^{\pm
v}_{\bi{d}}$ are the same for all sequences $\bi{\sigma}$ with the
same $\bi{d}$, which is the condition for lumping. The
mutation--selection model with Hopfield-type fitness is indeed
``lumpable'' with respect to the partition induced by the distance
vectors $\bi{d}$.

The mutation--selection process on the mutational distance space
$\mc{S}$ is described by the lumped time-evolution operator
$\bi{H}=\bi{R}+\bi{M}$ with lumped reproduction and mutation matrices
$\bi{R}$ and $\bi{M}$ of dimension $|\mc{S}|\times|\mc{S}|$. Whereas
the lumped reproduction matrix $\bi{R}$ is still diagonal and contains
the same entries as $\mc{R}$, i.e.\ $\bi{R}_{\bi{d}_{\bi{\sigma}}} =
\mc{R}_{\bi{\sigma}}$, the off-diagonal entries of the lumped mutation
matrix $\bi{M}$ are given by the cumulative mutation rates with
unchanged diagonal entries compared to $\mc{M}$, which still fulfil
the Markov property $\bi{M}_{\bi{dd}}=-\sum_{\bi{d}'\neq \bi{d}}
M_{\bi{d}'\bi{d}}$, and thus
\begin{equation}
\label{two-state lumped mutation matrix}
M_{\bi{d}'\bi{d}}=\begin{cases}
u^{+v}_{\bi{d}} & \text{if $\bi{d}'=\bi{d}+\bi{e}_v$,}\\
u^{-v}_{\bi{d}} & \text{if $\bi{d}'=\bi{d}-\bi{e}_v$,}\\
-\sum_v \left( u^{+v}_{\bi{d}} + u^{-v}_{\bi{d}} \right) = -\mu  
  & \text{if $\bi{d}'=\bi{d}$,}\\
0 & \mbox{otherwise,}
\end{cases}
\end{equation}
with the cumulative mutation rates from equation (\ref{cumulative
mutation rates}).  For a more general derivation of the lumped
reproduction and mutation matrices, see \citet{GG04a, Gar05}.

Note that the time-evolution operator $\bi{H}$ acting on $\mc{S}$
describes the evolution of a population under mutation and selection
determined by the evolution equation (\ref{evolution equation}), and
thus the theory developed in section \ref{The mutation--selection
model in sequence space} applies.

\section{The maximum principle}
\label{The maximum principle}

Although the lumping procedure reduces the number of types very
efficiently, the evaluation of the eigenvalues and eigenvectors of the
time-evolution operator $\bi{H}$ still remains a difficult problem for
many applications, due to the size of the eigenvalue problem. If one
is interested solely in the equilibrium behaviour of the system,
however, it is possible to determine the population mean fitness (at
least asymptotically for large sequence length $N$), given by the
leading eigenvalue of $\bi{H}$.  This can be done by a simple maximum
principle that can be derived from Rayleigh's general maximum
principle, which specifies that the leading eigenvalue
$\lambda_{\rm{max}}$ of an $n\times n$ matrix $\bi{H}$ can be obtained
via a maximisation over $\mathbb{R}^n$,
\begin{equation}
\label{Rayleigh's maximum principle}
\lambda_{\rm{max}} =\sup_{\bi{v}\in \mathbb{R}^n}
\frac{\bi{v}^T \bi{H v}}{\bi{v}^T \bi{v}} \,.
\end{equation}
The vector $\bi{v}$ for which the supremum is attained is the
eigenvector corresponding to the eigenvalue $\lambda_{\rm{max}}$.  The
simple maximum principle derived from this guarantees that the
population mean fitness $\bar{r}$ can be obtained by maximising a
function on the mutational distance space $\mc{S}$. It can be shown
that the maximiser itself is the ancestral mean mutational distance.

Such a maximum principle has first been derived by \citet{HRWB02} for
two-state sequences with permutation-invariant fitness. This has been
generalised to apply for four-state sequences with
permutation-invariant fitness by \citet{GG04a}, and subsequently the
restriction to permutation-invariant fitness function has been relaxed
by \citet{BBBK05}. The results from \citet{BBBK05} apply directly to
the Hopfield-type fitness treated here.

\subsection{Symmetrisation of $\bi{M}$}
\label{Symmetrisation of M}

Whereas the original mutation matrix $\mc{M}$ is symmetric, the lumped
mutation matrix $\bi{M}$ is no longer symmetric, as different numbers
of sequences are lumped into the different classes, therefore giving
rise to unequal cumulative forward and backward mutation rates.  To
derive the maximum principle, it is necessary to symmetrise the
mutation matrix $\bi{M}$.

$\bi{M}$ is reversible, i.e.,
\begin{equation}
\label{reversibility}
M_{\bi{dd}'} \pi_{\bi{d}'} = M_{\bi{d}'\bi{d}} \pi_{\bi{d}} \,,
\end{equation}
where $\bi{\pi} = (\pi_{\bi{d}})_{\bi{d}\in\mc{S}}$ is the stationary
distribution of the pure mutation process, which is given by the
equidistribution of types on $\mf{S}$, and thus given by the number of
sequences $n_{\bi{d}}$ that are lumped onto the same mutational
distance vector $\bi{d}$. The reversibility of $\bi{M}$ implies that
it can be symmetrised by the means of a diagonal transformation
$\bi{\Pi}:=\rm{diag} \{\pi_{\bi{d}}\}$, which yields the symmetrised
mutation matrix as
\begin{equation}
\label{symmetrised mutation matrix}
\bi{\widetilde{M}}=\bi{\Pi}^{-1/2}\bi{M\Pi}^{1/2}\,,
\end{equation}
with off-diagonal entries
\begin{equation}
\label{off-diagonal entries}
\widetilde{M}_{\bi{dd}'}=M_{\bi{dd}'}\sqrt{\pi_{\bi{d}'}/\pi_{\bi{d}}}
=M_{\bi{d}'\bi{d}}\sqrt{\pi_{\bi{d}}/\pi_{\bi{d}'}}
=\sqrt{M_{\bi{dd}'}M_{\bi{d}'\bi{d}}}
=\widetilde{M}_{\bi{d}'\bi{d}} \,.
\end{equation}
Using the cumulative mutation rates $u^{\pm v}_{\bi{d}}$, this reads
\begin{equation}
\label{symmetrised mutation rates}
\widetilde{M}_{\bi{d}'\bi{d}}
:=\widetilde{u}_{\bi{d}}^{\pm v}
=\widetilde{u}_{\bi{d}\pm\bi{e}_v}^{\mp v}
=\sqrt{u_{\bi{d}}^{\pm v}u_{\bi{d}\pm\bi{e}_v}^{\mp v}} 
\quad \mbox{if $\bi{d}'=\bi{d}\pm\bi{e}_v$ and $0$ otherwise,}
\end{equation}
as can be seen by using the explicit representation for the cumulative
mutation rates from equation (\ref{cumulative mutation rates}) with
the $n_{\bi{d}}$ from equation (\ref{nd}). Because $\bi{\Pi}$ is
diagonal, the diagonal entries of the mutation matrix are unchanged,
\begin{equation}
\widetilde{M}_{\bi{dd}} = M_{\bi{dd}} = -\mu
\end{equation}

As $\bi{R}$ is diagonal as well, it is not changed by the
transformation $\bi{\Pi}^{1/2}$, and thus this transformation also
symmetrises the time-evolution operator such that
\begin{equation}
\label{symmetrised time-evolution operator}
\bi{\widetilde{H}}
=\bi{\Pi}^{-1/2}\bi{H}\bi{\Pi}^{1/2} =\bi{R}+\bi{\widetilde{M}}
\end{equation}
is symmetric. 

Before symmetrisation, $\bi{H}$ was expressed as the sum of a Markov
generator $\bi{M}$ and a diagonal remainder $\bi{R}$. As the
transformation $\bi{\Pi}$ does not preserve the Markov property, this
is not the case for the symmetrised time-evolution operator in
(\ref{symmetrised time-evolution operator}). It is however useful to
split it up this way. To this end, let
\begin{equation}
\label{split symmetrised time-evolution operator}
\bi{\widetilde{H}}=\bi{E}+\bi{F}\,,
\end{equation}
where $\bi{F}$ is a (symmetric) Markov generator and $\bi{E}$ is the
(diagonal) remainder. The off-diagonal entries of $\bi{F}$ are given
by those of $\bi{\widetilde{M}}$ from equation (\ref{symmetrised
  mutation rates}), $F_{\bi{d}'\bi{d}}=\widetilde{M}_{\bi{d}'\bi{d}}$
for $\bi{d}'\neq\bi{d}$, whereas the Markov property requires as
diagonal entries
\begin{equation}
\label{diagonal entries of F}
F_{\bi{dd}} =-\sum_{\bi{d}'}F_{\bi{d}'\bi{d}}
=-\sum_v\left(\widetilde{u}_{\bi{d}}^{+v}
+\widetilde{u}_{\bi{d}}^{-v}\right) 
=-\sum_v\left(
       \sqrt{u^{+v}_{\bi{d}}u^{-v}_{\bi{d}+\bi{e}_v}}
      +\sqrt{u^{-v}_{\bi{d}}u^{+v}_{\bi{d}-\bi{e}_v}}
\right)
\end{equation}
The remainder $\bi{E}$ is given by
\begin{equation}
\label{remainder E}
\begin{split}
E_{\bi{d}} &=R_{\bi{d}}+\widetilde{M}_{\bi{dd}}-F_{\bi{dd}} \\
&=R_d-\sum_v\left(u^{+v}_{\bi{d}}+u^{-v}_{\bi{d}}\right)
 +\sum_v\left(
             \widetilde{u}_{\bi{d}}^{+v} + \widetilde{u}_{\bi{d}}^{-v}
        \right)
\end{split}
\end{equation}

\subsection{Continuum approach for the limit of infinite sequence 
  length}
\label{Continuum approach for the limit of infinite sequence length}

To deal with the case of infinite sequence length, it will prove
useful to use intensively scaled normalised versions of the
extensively scaled variables like the mutational distances. The
pattern in the Hopfield model, previously characterised by the {\em
  number} of sites $N_v$ in each subset $\Lambda_v$, will now be
described by the {\em fraction} of sites in $\Lambda_v$, given by
$X_v:=N_v/N$.  Similarly, we use normalised partial distances
\begin{equation}
x_v:=d_v/N_v \,,
\end{equation}
where $x_v\in[0,1]$, with the normalised mutational distance vector
$\bi{x}=(x_v)_{v=1 \ldots 2^p}$. The permutation-invariant model is
obtained in the case $p=0$.

For finite $N$, $\bi{x}$ takes rational values in a normalised
version of the mutational distance space $\frac{1}{N}\cdot\mc{S}
\subset\mc{D}$, where $\mc{D}$ is a compact domain in $\bb{R}^{2^p}$.
For $N\rightarrow\infty$, the vectors $\bi{x}$ become
dense in $\mc{D}$.

Assume that the entries of $\bi{\widetilde{H}}=\bi{E}+\bi{F}$ can be
approximated by functions $e$ and $f$ from $C^2_b(\mc{D},\bb{R})$,
i.e., twice continuously differentiable functions with bounded second
derivatives that map $\mc{D}$ onto $\bb{R}$ such that
\begin{align}
\label{assumption E}
E_{\bi{d}} &= e(\bi{x}_{\bi{d}})+\order \,,\\
\label{assumption F}
F_{\bi{d}'\bi{d}} &= f_{\bi{\Delta}}(\bi{x}_{\bi{d}}) +\order \,,
\end{align}
where $\bi{\Delta}=\bi{d}'-\bi{d}$ and the notation $\bi{x}_{\bi{d}}$
is used to emphasise that the normalised mutational distance $\bi{x}$
corresponding to a particular $\bi{d}$ is meant. In fact, assumption
(\ref{assumption F}) can readily be verified for the cumulative
mutation rates from equation (\ref{cumulative mutation rates}):

Let the functions $f_{\bi{\Delta}}(\bi{x})$ be 
\begin{equation}
\label{function f}
f_{\bi{\Delta}}(\bi{x})=
\begin{cases}
\widetilde{u}^v(\bi{x}_{\bi{d}}) & 
\text{if $\bi{\Delta}=\pm\bi{e}_v$,} \\
-2\sum\widetilde{u}^v(\bi{x}_{\bi{d}}) &
\text{if $\bi{\Delta}=0$,}\\
0 & \text{otherwise,}
\end{cases}
\end{equation}
where the functions $\widetilde{u}^v(\bi{x})$ are given by
\begin{equation}
\label{functions u tilde}
\widetilde{u}^v(\bi{x})
:=\sqrt{u^{+v}(\bi{x})u^{-v}(\bi{x})}\,,
\end{equation}
with the cumulative mutation rates $u^{\pm v}(\bi{x}_{\bi{d}}):=u^{\pm
v}_{\bi{d}}$, which read explicitly
\begin{equation}
\label{functions u explicitly}
u^{+v}(\bi{x}) = \mu X_v(1-x_v) 
\quad \mbox{and} \quad 
u^{-v}(\bi{x}) = \mu X_v x_v \,.
\end{equation}
Using a Taylor approximation, it can be shown that the differences
between the exact entries of $F$ as given in equations
(\ref{symmetrised mutation rates}) and (\ref{diagonal entries of F}),
and their approximations from equation (\ref{function f}), are indeed
of $\order$.

Assuming that also the reproduction rates $R_{\bi{d}}$ can be
approximated by a $C^2_b$ function $r(\bi{x})$ as
\begin{equation}
\label{assumption R}
R_{\bi{d}}=r(\bi{x}_{\bi{d}})+\order \,,
\end{equation}
then from equation (\ref{remainder E}), the matrix $\bi{E}$ is
approximated by
\begin{equation}
\label{function e}
e(\bi{x})=r(\bi{x})-\sum_v
\left( u^{+v}(\bi{x})+u^{-v}(\bi{x})
       -2\widetilde{u}^{v}(\bi{x}) \right)\,,
\end{equation}
fulfilling equation (\ref{assumption E}).
With the definition of the {\em mutational loss function} $g$ as
\begin{align}
\label{mutational loss function}
g(\bi{x}) 
:&=\sum_v \left( 
                 u^{+v}(\bi{x})+u^{-v}(\bi{x}) 
               - 2\widetilde{u}^v(\bi{x}) 
         \right)\\
&=\mu\sum_{v=1}^{2^p} X_v \left[ 1-2\sqrt{x_v(1-x_v)} \right] \,,
\end{align}
this reads explicitly
\begin{equation}
\label{function e with g}
e(\bi{x}) = r(\bi{x}) - g(\bi{x}) \,.
\end{equation}

We are now in a position to apply theorems 1 and 2 from
\citet{BBBK05}, which read for the mutation--selection model with
Hopfield-type fitness considered here

\stepcounter{thmno}
{\bf Theorem \arabic{thmno} 
(The maximum principle). } \\
{\em 
  (i) Assume that for the lumped mutation--selection model as set up 
  in section \ref{Lumping for the Hopfield-type fitness} it is 
  possible to approximate the reproduction rates $R_{\bi{d}}$ by a 
  $C^2_b$ function $r(\bi{x}_{\bi{d}})$ as specified in equation 
  (\ref{assumption R}), and that the $C^2_b$ function $e$ assumes its 
  global maximum in the interior} $\rm{int}(\mc{D})${\em. Then the 
  population mean fitness in equilibrium is given by
\begin{equation}
\label{maximum principle}
\bar{r} = \sup_{\bi{x}\in\mc{D}}
\left[ r(\bi{x})-g(\bi{x}) \right] +\order \,.
\end{equation}
(ii) 
  Assume furthermore that $e$ assumes its maximum at a unique point
  $\bi{x}^*$ in} $\rm{int}(\mc{D})${\em, and that the Hessian of $e$ 
  at $\bi{x}^*$ is negative definite. Then in the limit of
  $N\rightarrow\infty$, the maximiser $\bi{x}^*$ is given by the mean
  ancestral mutational distance $\bi{\hat{x}}$, and in particular
\begin{equation}
\label{ancestral maximum principle}
\bar{r}=r(\bi{\hat{x}})-g(\bi{\hat{x}}) \,.
\end{equation}
} 
For a proof the reader is referred to \citet{BBBK05}.

\section{Error thresholds}
\label{Error thresholds}

The maximum principle (\ref{maximum principle}) and (\ref{ancestral
  maximum principle}) is a powerful tool to calculate the population
mean fitness $\bar{r}$ in equilibrium for arbitrary fitness
functions of the permutation-invariant or Hopfield type, for any
range of mutation rates. Also, the ancestral mean genotype
$\bi{\hat{x}}$ is available. The general method to identify
$\bar{r}$ and $\bi{\hat{x}}$ is to consider the partial derivatives
of $r-g$ with respect to the components $x_v$ of the mutational
distance $\bi{x}$. A necessary condition for the function $r-g$ to
have a maximum at a value $\bi{x}^*$ is that its derivatives at this
$\bi{x}^*$ vanish,
\begin{equation}
\label{error threshold general method}
\frac{\partial}{\partial x_{v,k}} 
\left[r(\bi{x})-g(\bi{x})\right]_{\bi{x}=\bi{x}^*} = 0 
\quad\forall\, v,k \;.
\end{equation}
The global maximum of the function $r-g$ must lie on one of the points
$\bi{x}^*$ that fulfil equation (\ref{error threshold general method})
or on the border of the mutational distance space. Thus by comparing
the values of $r-g$ on these possible points, the global maximum can
be identified.

Apart from the general possibility to investigate the
dependence of the population mean fitness $\bar{r}$ on the mutation
rate $\mu$, this yields the opportunity to investigate the phenomenon
of the {\em error threshold}, which has interested scientists ever
since it was first conceived by \citet{Eig71}.

The phenomenon of the error threshold can be described as the
existence of a critical mutation rate, below which the equilibrium
population is well localised in sequence space, whereas for mutation
rates above the critical mutation rate, the equilibrium population is
more delocalised, with a sharp transition between the two phases. 

One problem is that there is no generally accepted definition of an
error threshold. The criterion used in the original quasispecies model
\citep{Eig71} is the disappearance of the wild-type from the
population, which under the single peaked landscape used there goes in
line with the complete delocalisation of the population in sequence
space. However, these two effects do not necessarily coincide for
other fitness landscapes.

The definition of the error threshold that shall be used here is
equivalent to the definition of a phase transition in physics,
differentiating between first and second order transitions as follows:

{\bf Definition (First and second order error threshold).}\\
  (i) A first order error threshold exists at a critical mutation rate
  $\mu_c$, if the ancestral mean mutational distance as a function of
  the mutation rate $\bi{\hat{x}}(\mu)$ shows a discontinuity at this
  $\mu_c$, which is also reflected by a kink in the population mean
  fitness $\bar{r}(\mu)$.\\
  (ii) A second order error threshold exists at a critical mutation
  rate $\mu_c$, if the ancestral mean mutational distance is
  continuous, but its derivative with respect to the mutation rate
  $\left[\frac{d \bi{\hat{x}}} {d \mu}\right]_{\mu\rightarrow\mu_c}$
  is discontinuous at this mutation rate $\mu_c$.

In the examples shown later in this thesis, the second order error
threshold always show an infinite derivative at the critical mutation
rates.  Note that, like phase transitions in physics, these
definitions of the error thresholds apply in the strict sense only to
a system with infinite sequence length ($N\rightarrow\infty$), for
finite sequence lengths, the thresholds are smoothed out due to the
lack of non-analyticities.

\cite{HRWB02} gave a finer classification of different error threshold
phenomena. The first order error threshold they called ``fitness
threshold''. Here, this term shall include also the second order error
threshold, making all error thresholds as defined above fitness
thresholds. Furthermore, the concept of the ``degradation threshold''
was introduced:

{\bf Definition (Degradation threshold).}  \\
  A degradation threshold is an error threshold of first or second 
  order, where the population distribution beyond the critical 
  mutation rate $\mu_c$ is given by the equidistribution in sequence 
  space $\mf{S}$.  

Thus here the degradation threshold is a special case of a fitness
threshold, going in line with the complete delocalisation of the
population in sequence space. Note that in the limit of infinite
sequence length ($N\rightarrow\infty$), for which the error threshold
definitions apply exactly, this equidistribution is reached
immediately above $\mu_c$, and beyond the threshold the population is
insensitive to any further increase in mutation rates.  In the case of
finite sequence lengths, where the thresholds are smoothed out, the
equidistribution is of course only reached asymptotically.

The original error threshold was observed for the single peaked
fitness landscape, where a single sequence is attributed a high
fitness value, all other sequences are equally disadvantageous
\citep[for a review, see][]{EMS89}. This is clearly an
oversimplification and should not be regarded as anything but a toy
model. Other fitness landscapes that have been investigated comprise,
in the permutation-invariant case, linear and quadratic fitness
functions, general functions showing epistasis, and as examples
lacking permutation-invariance the Onsager landscape \citep{BBW97,
BW01}, which has nearest neighbour interactions within the sequence,
as well as various spin glass landscapes like the Hopfield landscape
\citep{Leu87, Tar92}, the Sherrington-Kirkpatrick spin glass
\citep{BS93b}, the NK spin glass \citep{CAW02}, and the random energy
model \citep{FPS93, FP97}, assigning random fitness values to each
sequence.

One fitness landscape where an analytical solution can be obtained is
the linear fitness \citep[cf.][]{Rum87, Hig94a, BBW97}. Note that
this corresponds to a multiplicative landscape in a set-up using
discrete time. For a linear fitness function, there is no error
threshold, but the population changes smoothly from localised to
delocalised with an increasing mutation rate.

For quadratic fitness functions, error thresholds only exist for
antagonistic epistasis; they are absent for quadratic fitness
functions with synergistic epistasis \citep{BBW97, HWB01, GG04b}.
These results go in line with those for general epistatic fitness
functions \citep{Wie97}. Studies using non-permutation-invariant
fitness functions generally report the presence of error thresholds.

Of course the discussion of the error threshold phenomenon is academic
if the threshold is an artifact of the model rather than a real
biological phenomenon. This issue has been subject to numerous
debates, especially because it has first been predicted by a model
using the over-simplistic single peaked landscape. However, over the
years biologists have accumulated evidence that particularly RNA
viruses naturally thrive at very high mutation rates \citep{DH88,
  EB88}, of the order of $10^{-4}$ to $10^{-5}$ per base per
replication \citep{DES+96}, corresponding to a genomic mutation rate
of about 0.1 to 10 mutations per replication \citep{DH97}, and a
number of studies have reported that populations of RNA viruses only
survive a moderate increase of their mutation rate, whereas if the
mutation rate is increased further, the populations become extinct
\citep{HDdlTS90, LEK+99, SDLD00, CCA01}, for reviews see
\citet{DES+96, DH97}. This corresponds to the population being pushed
beyond the error threshold.  It has been suggested to use the error
threshold for anti-viral therapies \citep{Eig93}, and in fact, recent
experimental results indicate that this is the mechanism via which the
broad-spectrum anti-viral drug ribavirin works \citep{CCA01}.  This
clearly warrants some further investigation of the error threshold
phenomenon, which shall be done in the remainder of this section.

\subsection{Different Hopfield-type fitness functions}

The original Hopfield fitness as introduced by \citet{Hop82} is a
quadratic function of the specific distances and reads
\begin{equation}
\label{original Hopfield fitness}
r_{\bi{\sigma}} = 
-\sum_{q=0}^p y^q + \sum_{q=0}^p \left(y^q\right)^2 \,,
\end{equation}
using normalised specific distances $y^q:=w^q/N$, similarly to the
normalised mutational distances $\bi{x}$.  The statistical properties
of this landscape have been studied in detail
\citep{AGS85a,AGS85b,Tal03}: In the thermodynamic limit $N \rightarrow
\infty$, there are $2(p+1)$ global maxima that are associated with the
patterns $(\bi{\xi}^q)$ and their complements $(\bi{1}-\bi{\xi}^q)$.
In addition to that, the number of local maxima and saddle points
grows exponentially with the number patterns $p+1$, hence the
ruggedness of the fitness landscape can be tuned by the number of
patterns.

Most works that have studied a Hopfield-type fitness used the original
Hopfield model, a generalisation was however treated by \citet{Pel02},
using a Hopfield-type truncation selection with two patterns. Thus it
might be interesting and instructive to investigate the threshold
behaviour of different kinds of Hopfield-type fitness functions.

Applying criteria for the existence of error thresholds that have been
obtained by \citet{HRWB02} for permutation-invariant fitness functions
to the case of a Hopfield-type fitness, it can be shown that for
linear Hopfield-type fitness functions there are no error thresholds,
which is a new result, considering that for all previously
investigated Hopfield-type fitness functions, the existence of error
thresholds was reported.

The next step towards more complex fitness functions is to consider
quadratic fitness functions, generalising the original
Hopfield-fitness, which is a particular example for a quadratic
function. Here, the analysis shall be restricted to a symmetry with
respect to the normalised specific distances $y^q$ to the patterns
$\bi{\xi}^q$, such that
\begin{equation}
\label{quadratic symmetric Hopfield-type fitness}
r = c \sum_{q=0}^p y^q \pm \sum_{q=0}^p \left( y^q \right)^2 \,.
\end{equation}
The parameter $c$ tunes the linear in relation to the quadratic term,
and the sign of the quadratic term determines the {\em epistasis}, a
measure for the strength of interaction between sites. For a positive
quadratic term epistasis is said to be negative or antagonistic,
whereas for a negative quadratic term one speaks of positive or
synergistic epistasis.  The case $c=-1$ combined with a positive
quadratic term (i.e., negative epistasis) yields the original Hopfield
fitness.

\subsection{Quadratic symmetric Hopfield-type fitness with two 
  patterns}

In the case of two patterns, $p=1$, the first pattern can be chosen
without loss of generality as $\bi{\xi}^0=00\ldots0$, such that there
is only one pattern to be chosen, usually randomly. The matrix
$\bi{\rho}$ containing the possible types of sites is given by
\begin{equation}
\label{two-state two-pattern rho}
\bi{\rho}=\left(\begin{array}{cc} 0&0\\0&1 \end{array}\right)\;,
\end{equation}
and thus the index set of sites is partitioned into two subsets,
$\Lambda=\Lambda_1\cup\Lambda_2$, where $\Lambda_1$ contains all sites
at which both patterns have entry $0$, whereas $\Lambda_2$ corresponds
to the sites where the two patterns have entries $0$ and $1$,
respectively. The only quantities characterising the patterns are now
the fractions of sites in each partition, $X_1=N_1/N$
and $X_2=N_2/N=1-X_1$. Thus the pattern can be characterised by a
single parameter, $X_1$.

Each sequence is characterised with respect to the pattern by the
partial Hamming distances to pattern $\bi{\xi}^0$ (in normalised
form), $x_1$ and $x_2$. These vary from $0$ (all entries $0$ in
$\Lambda_v$) to $1$ (all entries $1$ in $\Lambda_v$), completely
independently from each other. The specific distances with respect to
the patterns are linear combinations of the $x_v$ and given in
normalised form by
\begin{align}
\label{two-state two-pattern specific distances}
y^0=w^0/N &=X_1 x_1+X_2 x_2 \,, \notag\\
y^1=w^1/N &=X_1 x_1+X_2 (1-x_2)\;.
\end{align}
The Hopfield-type fitness is defined as an arbitrary function of these
patterns, $r=f(y^0, y^1)$. Due to the small number of variables, for
the case of two patterns, a lot can be done by analytical treatment.

For the quadratic symmetric Hopfield-type fitness (\ref{quadratic
  symmetric Hopfield-type fitness}) with positive epistasis (i.e.,
negative sign of the quadratic term) and $c=1$, there are no phase 
transitions, going in line with the results for permutation-invariant 
fitness functions, but different from other results for Hopfield-type 
fitness functions. As an example for negative epistasis, consider 
first the original Hopfield fitness (\ref{original Hopfield fitness}).

\subsubsection{The original Hopfield fitness with two patterns}

\begin{figure}
\centerline{
\includegraphics[width=0.45\textwidth]
{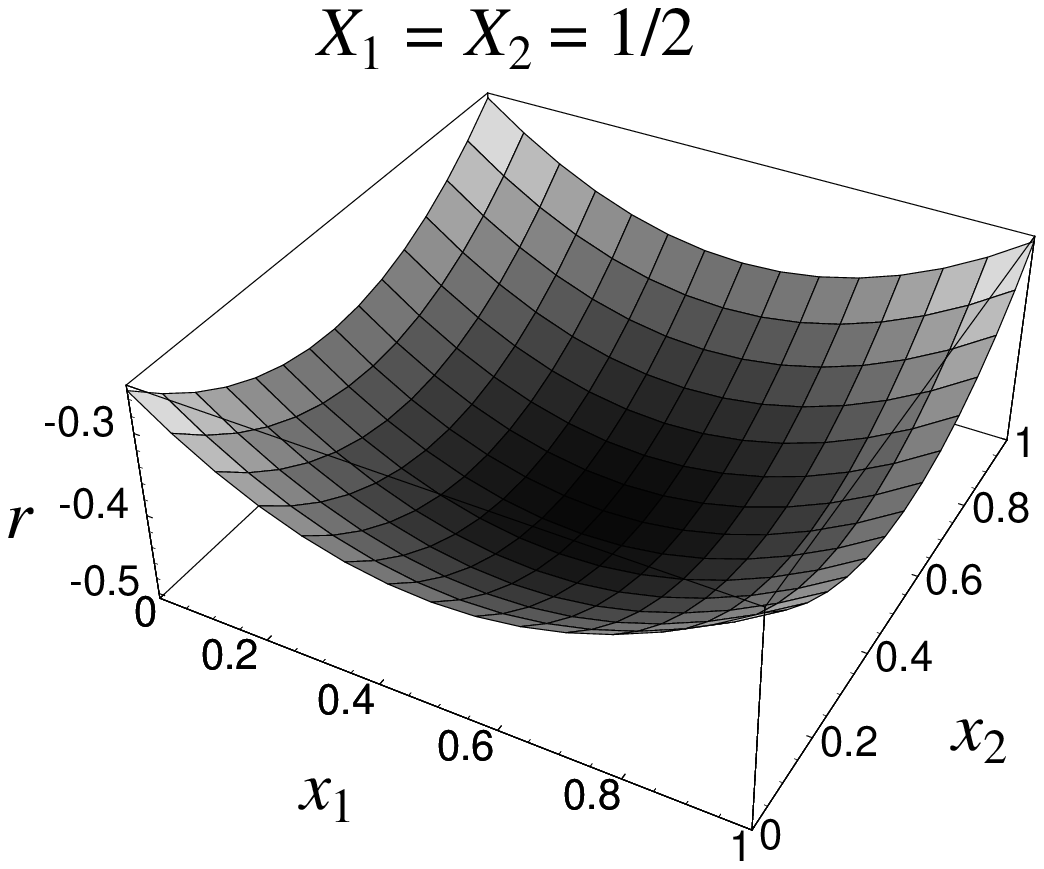}
\hspace{0.05\textwidth}
\includegraphics[width=0.45\textwidth]
{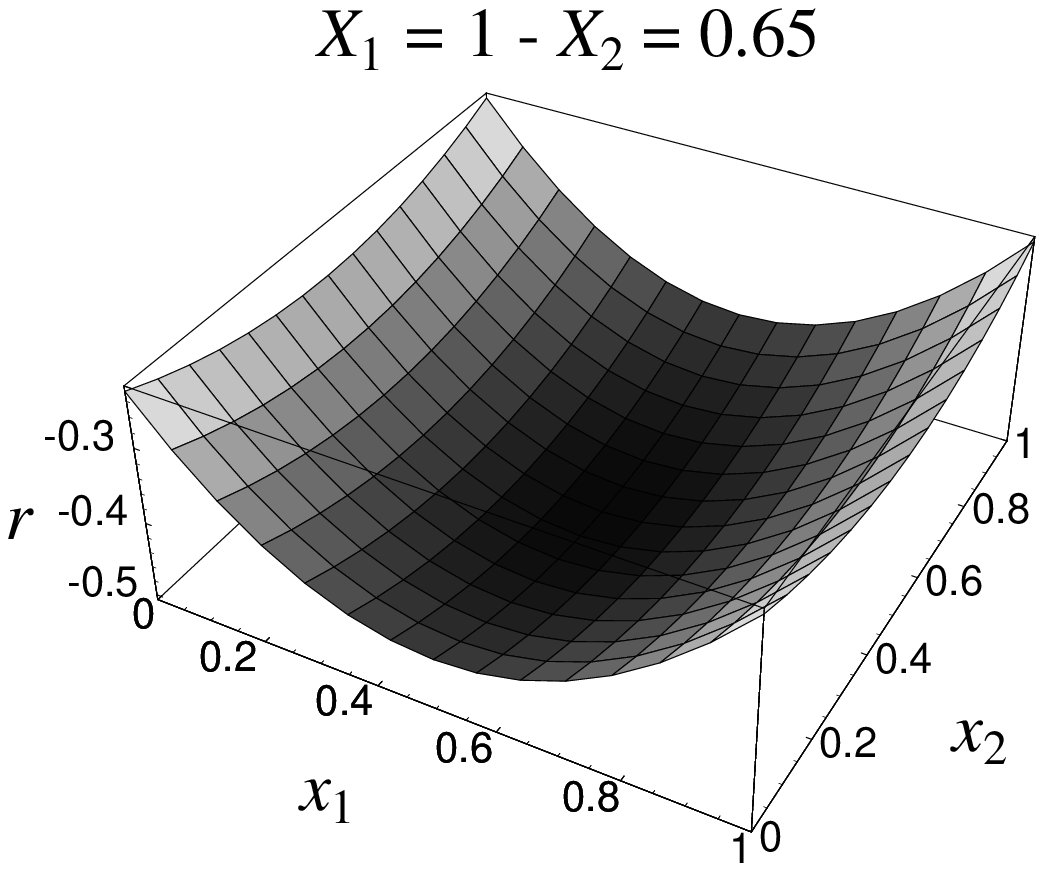}
}
\caption[Original Hopfield fitness for two patterns.]
{\label{Fig 6.14}Original Hopfield fitness in the case of two 
  patterns with different $X_1, X_2$.}
\end{figure}

Since for two patterns there are only two variables, it is possible to
visualise the fitness landscape in this case. Figure \ref{Fig 6.14}
shows the original Hopfield fitness (\ref{original Hopfield fitness})
for the cases $X_1=X_2=1/2$ and $X_1=1-X_2=0.65$.
In the corners of the mutational distance space $\mc{S}$, one can see
the four degenerate maxima. 

\begin{figure}
\centerline{
\includegraphics[width=\textwidth]
{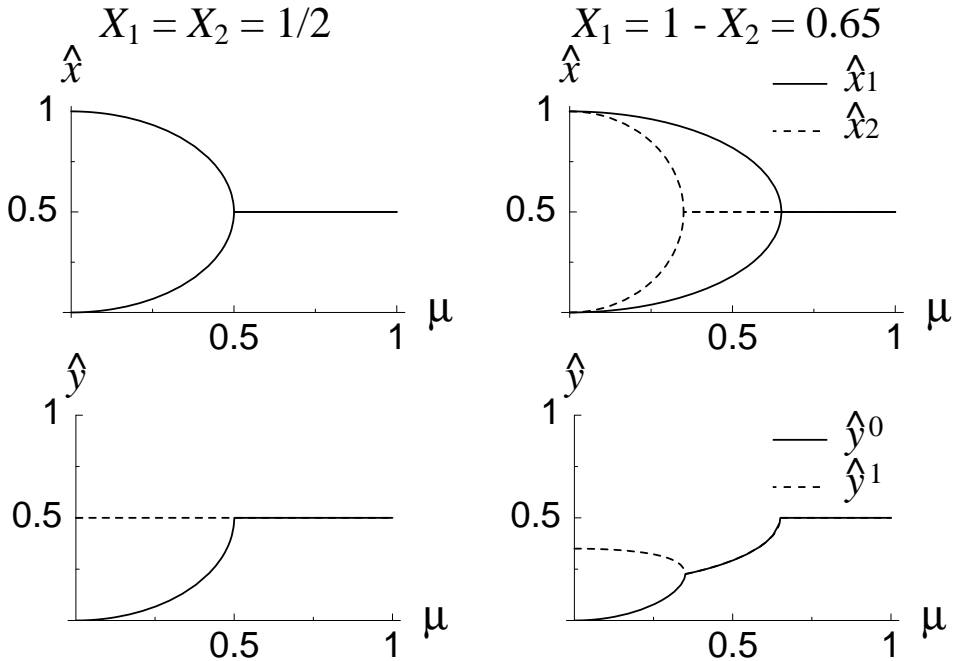}
}
\caption[Ancestral mean partial and specific distances, $\hat{x}_v$ 
         and $\hat{y}^q$, for the original Hopfield fitness with two 
         patterns.]
{\label{Fig 6.15}Ancestral mean partial distances $\hat{x}_v$ (top)
  and specific distances $\hat{y}^q$ (bottom) depending on mutation
  rates. The original Hopfield fitness (\ref{original Hopfield
  fitness}) for two patterns has been used. Results correspond to
  uncorrelated patterns ($X_1=X_2=1/2$, left) and correlated patterns
  with with $X_1=1-X_2=0.65$ (right).}
\end{figure}

The ancestral mean partial distances $\hat{x}_v$, at which the maxima
of $r-g$ are positioned, are obtained by considering the derivatives
of $r-g$. They are given by
\begin{equation}
\label{two-state two-pattern analytical solution}
\hat{x}_v =
\begin{cases}
\frac{1}{2}\pm\frac{1}{2X_v}\sqrt{X_v^2-\mu^2} 
& \text{for $\mu\le X_v$,}\\
\frac{1}{2} & \text{for $\mu\ge X_v$.}
\end{cases}
\end{equation}
For the case of $X_1=X_2=1/2$, which corresponds to two completely
uncorrelated patterns, the ancestral mean partial distances
$\hat{x}_v$ are shown in figure \ref{Fig 6.15} on the left, alongside
the ancestral mean specific distances $\hat{y}^q=y^q(\hat{x}_v)$.
For low mutation rates, there are two possible solutions for each of
the ancestral mean partial distances $\hat{x}_v$, and as the maxima
are degenerate, in equilibrium, the population will be centred equally
around all of them. However, in the approach to equilibrium, the
population might well be predominantly concentrated around one of
them, depending on initial conditions. The specific distances $y^q$
that are shown correspond to the combination of $\hat{x}_1$ and
$\hat{x}_2$, where both are given by the lower branch. Other
combinations yield similar results. For high mutation rates, the
population is in the mutation equilibrium with $\hat{x}_1 =\hat{x}_2
=1/2$, forming a disordered phase. In the limit of low mutation rates,
$\mu\rightarrow0$, the population is always in the vicinity of one of
the patterns (or its complement), such that one of the
$\hat{y}^q\approx0\;(1)$, which is completely random with respect to
the other pattern, and thus the other $\hat{y}^q=1/2$. This is the
ordered phase. At the critical mutation rate $\mu_c=1/2$, there is a
second order phase transition between these two phases, which is a
fitness as well as a degradation threshold, corresponding to the
infinite derivative of both $\hat{x}_v$ at this mutation rate. As the
specific distances $y^q$ are simply superpositions of the partial
distances $x_v$, the phase transitions are also visible in the $y^q$.

In the correlated case $X_1\neq X_2$ (figure \ref{Fig 6.15}, right),
two second order transitions can be identified. At $\mu=X_1$,
$\hat{x}_1$ has a phase transition, whereas at $\mu=X_2$, $\hat{x}_2$
has a phase transition. The threshold occurring at the lower mutation
rate is only a fitness threshold, whereas the one happening at the
higher mutation rate is both a fitness and degradation threshold,
leading to a totally random population. For $0\le\mu\le\min(X_1,X_2)$,
the population is in an ordered phase, for $\min(X_1,X_2) \le \mu \le
\max(X_1,X_2)$, it is in a partially ordered phase, which is ordered
with respect to one of the variables, but random with respect to the
other. Finally, for $\mu\ge\max(X_1,X_2)$, the population is the
equidistribution in sequence space. Here again, for low mutation rates
the population is close to one of the patterns, but due to the
correlation in the chosen patterns, this leads to a non-random overlap
with the other pattern. In the uncorrelated case with $X_1=X_2=1/2$,
the two error thresholds coincide, and the partially ordered phase
vanishes.

\subsubsection{Deviations from the original Hopfield model}

Now turn to the question how these phase transitions depend on the
particular degeneracy of the fitness functions and consider the
quadratic fitness function (\ref{quadratic symmetric Hopfield-type
fitness}) with negative epistasis (i.e., positive quadratic term) for
values of $c\neq-1$.

\begin{figure}
\centerline{
\includegraphics[width=0.47\textwidth]
{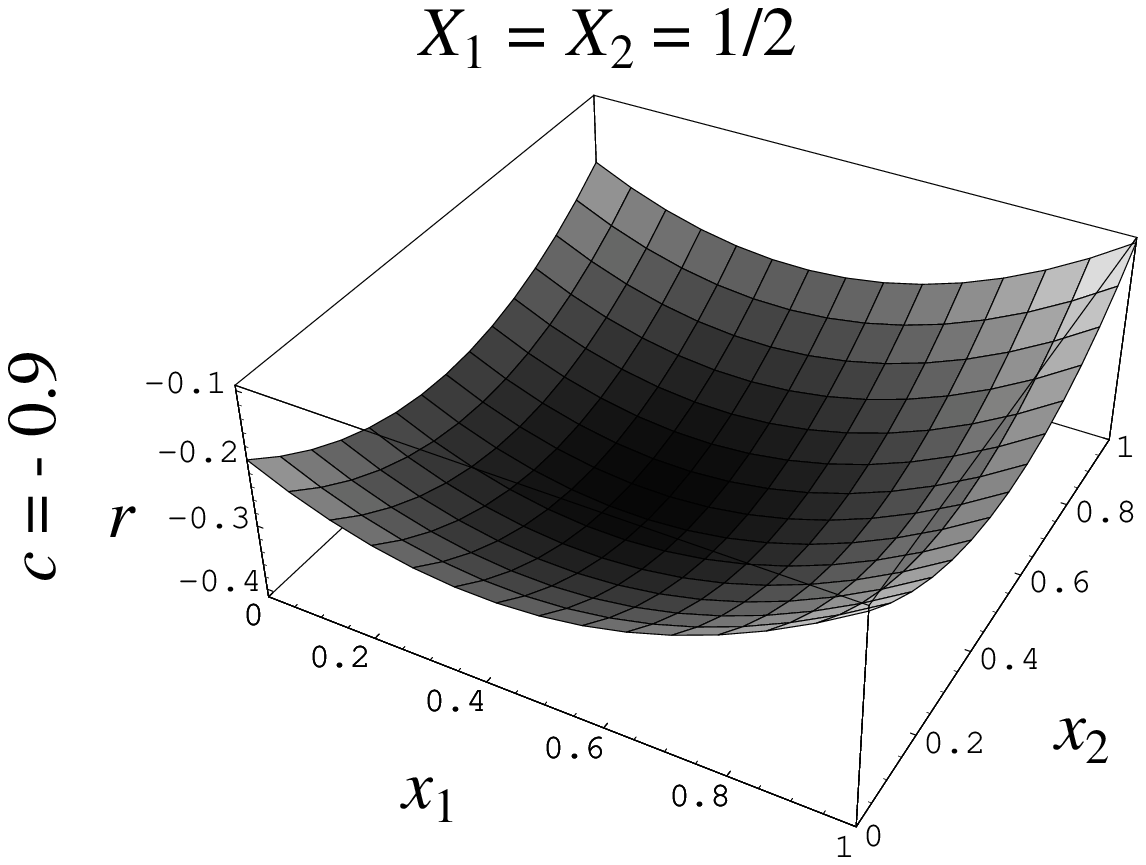}
\hspace{0.05\textwidth}
\includegraphics[width=0.43\textwidth]
{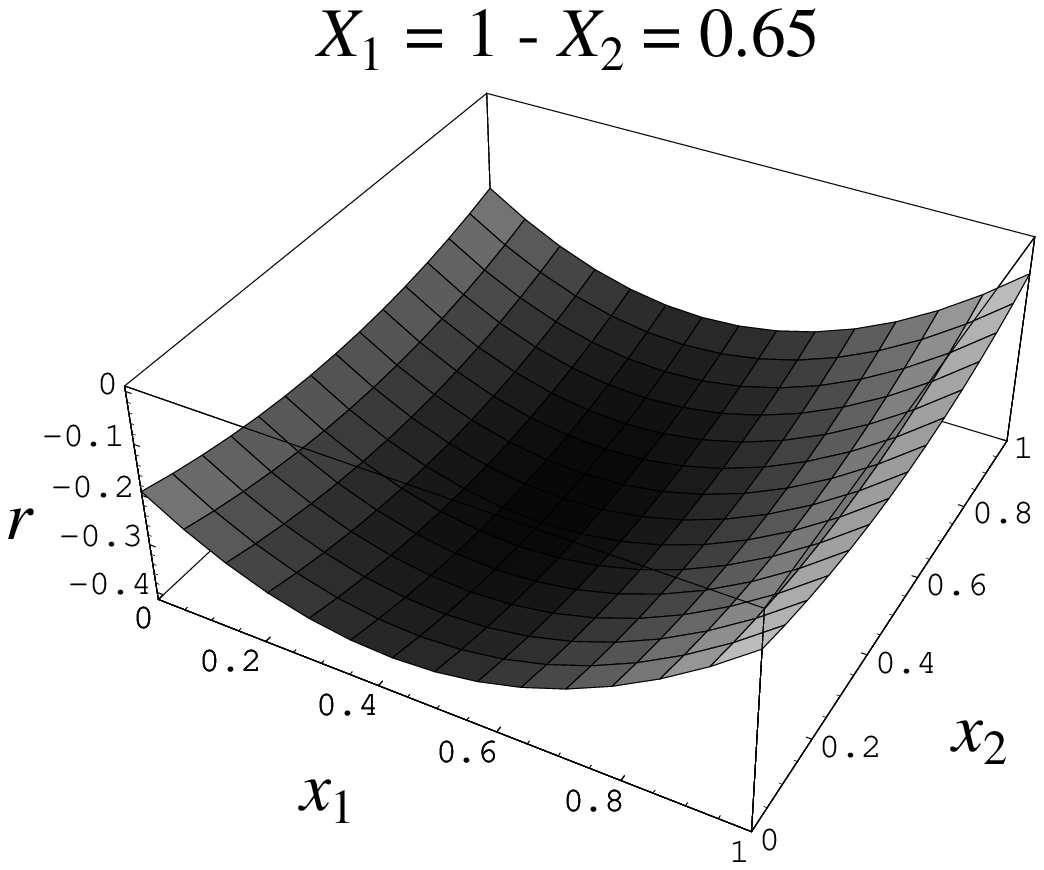}
}
\centerline{
\includegraphics[width=0.47\textwidth]
{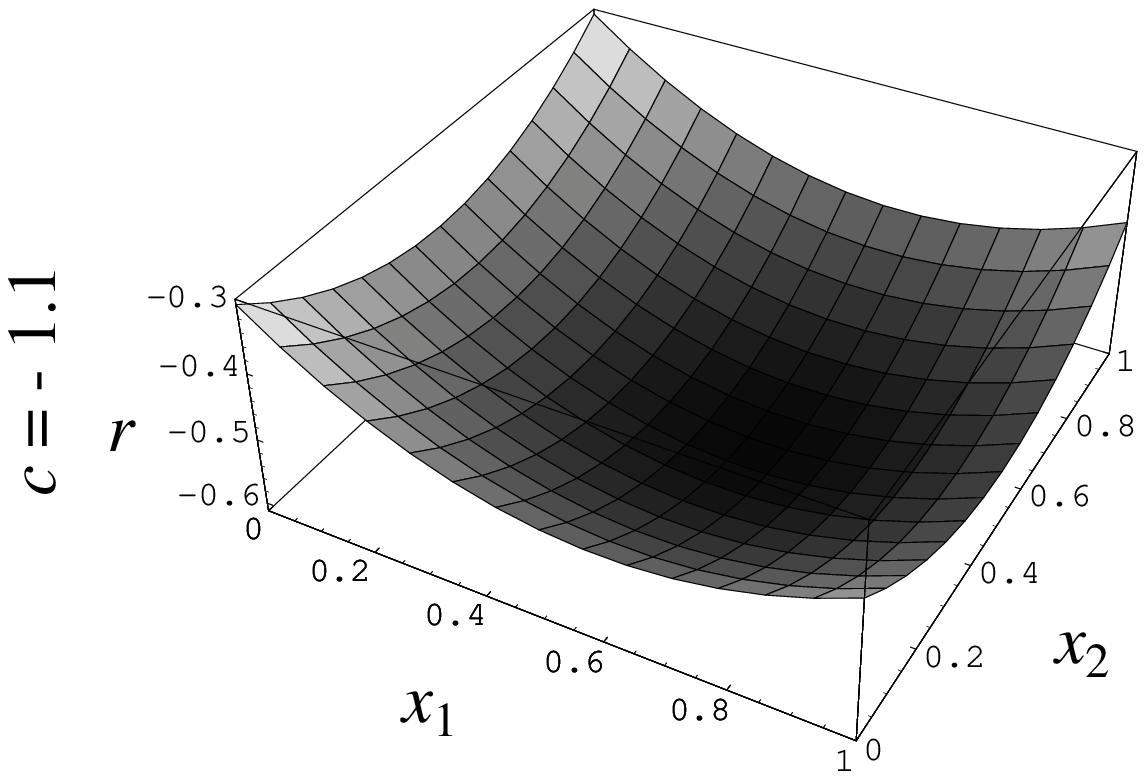}
\hspace{0.05\textwidth}
\includegraphics[width=0.43\textwidth]
{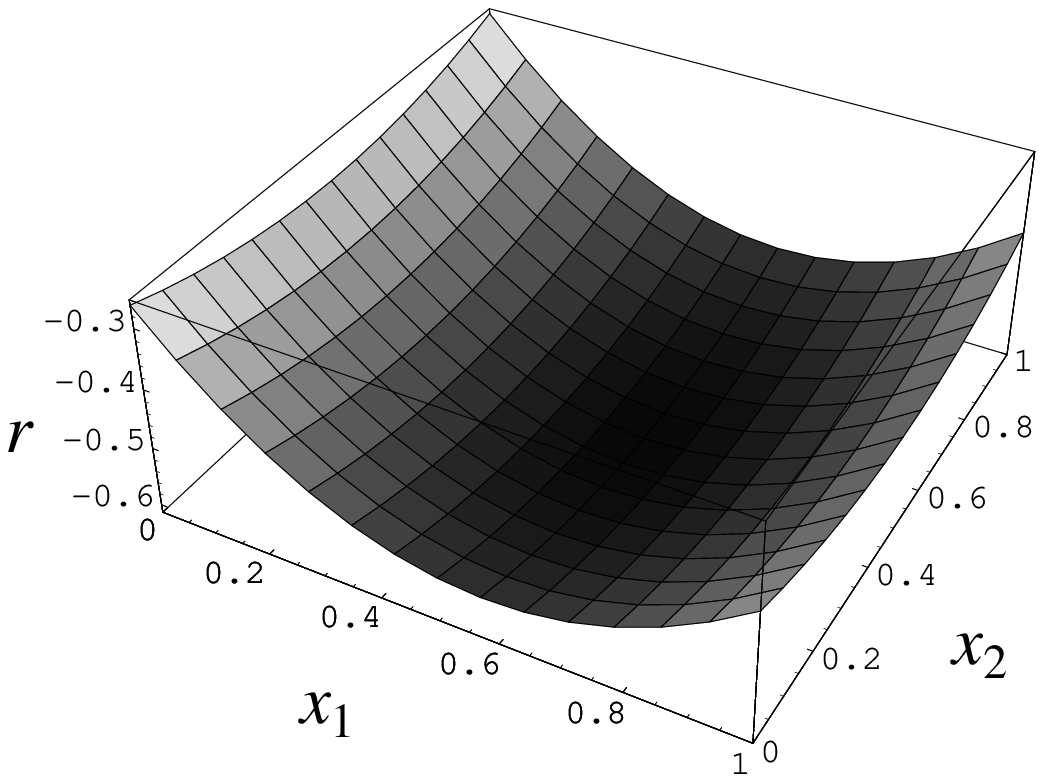}
}
\caption[Quadratic Hopfield-type fitness functions with negative 
         epistasis.]
{\label{Fig 6.16}Quadratic Hopfield-type fitness functions
  (\ref{quadratic symmetric Hopfield-type fitness}) with
  negative epistasis and $c=-0.9$ (top) and $c=-1.1$ (bottom) for an
  uncorrelated pattern $X_1=X_2=1/2$ (left) and a correlated pattern
  with $X_1=1-X_2=0.65$ (right).}
\end{figure}

Figure \ref{Fig 6.16} shows the fitness landscapes for values of
$c=-0.9$ and $c=-1.1$ for uncorrelated patterns $X_1=X_2=1/2$ and
correlated patterns with $X_1=1-X_2=0.65$.
As the pictures indicate, the fitness functions (and thus the
behaviour of the system) with the same $|c+1|$ are related by symmetry
under $x_1\rightarrow 1-x_1$ (apart from a constant term, which does
not influence the dynamics). Note that in $x_2$-direction, the fitness
function is independent of $c$. This is because in the sum of the
specific distances, the term with $x_2$ cancels out, which happens
generally in the case of an even number of patterns (i.e., odd $p$)
for $\left(\atop{p}{(p+1)/2}\right)$ different variables.

\begin{figure}
\includegraphics[width=\textwidth]{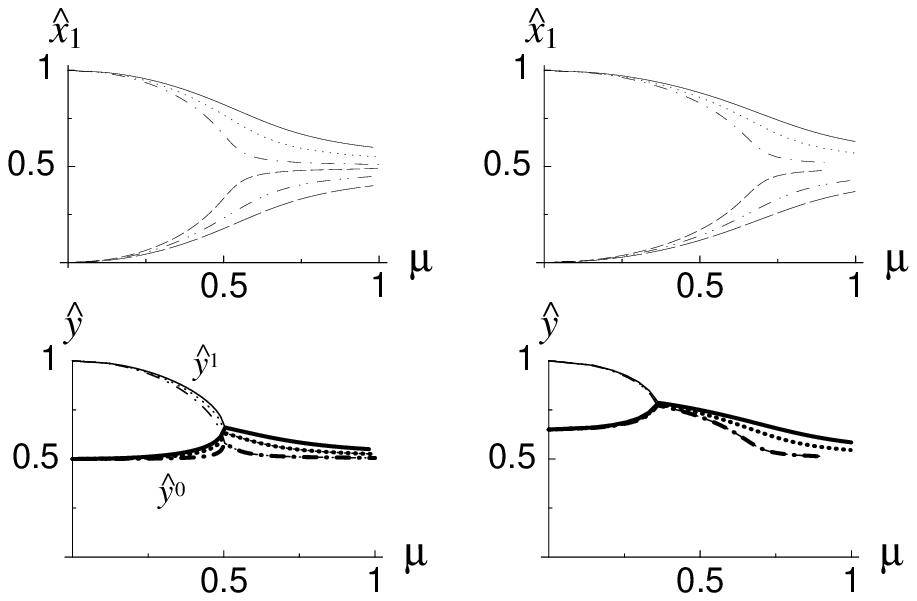}
\caption[Ancestral mean partial and specific distances, $\hat{x}_1$, 
         $\hat{y}^0$ and $\hat{y}^1$, for quadratic Hopfield-type 
         fitness with two patterns.]
{\label{Fig 6.17}Ancestral mean partial distance $\hat{x}_1$ (top) and
  specific distances $\hat{y}^0$ (thick lines, bottom) and $\hat{y}^1$
  (thin lines, bottom) depending on mutation rates. The quadratic
  Hopfield-type fitness (\ref{quadratic symmetric Hopfield-type
  fitness}) with negative epistasis for two patterns and different
  values of $c$ has been used. Results correspond to uncorrelated
  patterns $X_1=X_2=1/2$ (left) and correlated patterns with
  $X_1=1-X_2=0.65$ (right). Data are shown for parameter values of
  $c=-0.9, -0.95, -0.99, -1.01, -1.05, -1.1$ (top to bottom). For
  clarity, only specific distances $\hat{y}^q$ corresponding to $c>-1$
  are shown. }
\end{figure}

Because in $x_2$-direction the fitness is independent of $c$, the
solution for the ancestral mean mutational distance $\hat{x}_2$ is
identical with the solution for the original Hopfield fitness as given
in equation (\ref{two-state two-pattern analytical solution}). So for
all values of $c$, the phase transition with respect to $x_2$ happens
at $\mu=X_2$. For $x_1$, the solution becomes more complicated, but
the inverse function is simpler. It is given by
\begin{equation}
\mu = \frac
{2\left[1+c+X_1(2\hat{x}_1-1)\right]\sqrt{\hat{x}_1(1-\hat{x}_1)}}
{2\hat{x}_1-1} \,.
\end{equation}
The dependence of $\hat{x}_1$ on the mutation rate is shown in figure
\ref{Fig 6.17} (top).
For $c\neq-1$, the second order phase transition is smoothed out and
thus vanishes. Note that the ambiguity in the solutions that exists in
the case $c=-1$ (cf.\ figure \ref{Fig 6.15}), does not exist here, due
to the lacking degeneracy of the maxima of the fitness function at
$x_1=0$ and $x_1=1$ (cf.\ figure \ref{Fig 6.16}). At the bottom,
figure \ref{Fig 6.17} shows the specific distances $\hat{y}^q$, using
the lower branch of the solution for $\hat{x}_2$ (as shown in figure
\ref{Fig 6.15}), which show the second order transition in
$\hat{x}_2$, a fitness threshold. With this combination of solutions,
for low mutation rates the population is centred around the sequence
complementary to pattern $\bi{\xi}^1$. The general picture for the
uncorrelated ($X_1 = 1/2$) and correlated ($X_1 \neq 1/2$) choice of
patterns is very similar, apart from issues like the exact location of
the thresholds.

\subsection{Quadratic symmetric Hopfield-type fitness with three 
   and more patterns}
\label{The case of three patterns}

The behaviour of the system with quadratic symmetric Hopfield-type
fitness has been investigated for three, four and five
patterns. However, due to the complexity of the analysis for a higher
number of patterns, the focus is on the case of three patterns.

For three patterns, the matrix $\bi{\rho}$ reads
\begin{equation}
\label{two-state three-pattern rho}
\left(\begin{array}{cccc}
0&0&0&0\\
0&0&1&1\\
0&1&0&1
\end{array}\right)\;,
\end{equation}
thus there are four $X_v$ describing the patterns $\bi{\xi}$,
fulfilling $\sum_{v=1}^4 X_v =1$, and four variables $x_v\in[0,1]$,
describing each sequence. The specific distances with respect to
pattern $\bi{\xi}^q$ are given by
\begin{align}
\label{two-state three-pattern specific distances}
y^0 &= X_1 x_1 + X_2 x_2 + X_3 x_3 + X_4 x_4 \,, \\
y^1 &= X_1 x_1 + X_2 x_2 + X_3 (1-x_3) + X_4 (1-x_4) \,, \\
y^2 &= X_1 x_1 + X_2 (1-x_2) + X_3 x_3 + X_4 (1-x_4) \;.
\end{align}

Similarly to the case of two patterns, the original Hopfield fitness
(\ref{original Hopfield fitness}) shall be considered first, and then
variations (\ref{quadratic symmetric Hopfield-type fitness}) with
negative epistasis, but $c\neq-1$. The investigation is done by means
of numerical calculations.

\subsubsection{Variations of the pattern}

For the Hopfield-type fitness, the number of classes in the lumped
system is given by $|\mc{S}|= \prod_{v=1}^{2^p}(N_v+1)$.  Now consider
the case where the patterns are chosen randomly with equal probability
for each letter at each site. This results in a multinomial
probability distribution for the number of sites $N_v$ in each subset
\citep{Whi76},
\begin{equation}
\label{random pattern distribution}
\mc{P}(\{N_1,\ldots,N_n\})=\frac{N!}{n^N\prod_{v=1}^n N_v!},
\end{equation}
where $n=2^p$. The means are given by $\bar{N_v}=N/n$ and the variance
$\sigma_v^2=N(n-1)/n^2$ such that $N_v=N/n+\mc{O}(\sqrt{N})$. 

If the patterns $\bi{\xi}^q$ for $q=1,\ldots,p$ are chosen randomly
(remember $\bi{\xi}^0=00\ldots0$), the $X_v$ are thus given by $X_v
=2^{-p} +\mc{O}\left(\frac{1}{\sqrt{N}}\right)$.  So to mimic the
case of an infinite sequence length, in which the maximum principle is
exact, one has to assume uncorrelated patterns with $X_v=2^{-p}$
for all $v=1,\ldots,2^{p}$. However, the maximum principle can also
be used to investigate the case of finite sequence length by
simulating the finite sequence length through choosing pattern
distributions that do not follow exactly the infinite distribution
$X_v=2^{-p}$, but vary around this mean value with a variance
according to the sequence length to be simulated.  Practically,
patterns corresponding to finite sequence length $N$ have been
obtained by choosing for the $3N$ sites entries $0$ or $1$ with
probability 1/2 at each site, and counting the number of sites $N_v$
in each class $\Lambda_v$, similarly to the example pattern given in
equation (\ref{example pattern}). Thus although the patterns were
chosen randomly, they are correlated due to the finite sequence
length.

As shall be seen in the following, these correlations do account for
some interesting additional features. However, focus first on the case
of a genuinely infinite sequence length with $X_v=2^{-p}$ for all
$v$.

\subsubsection{The case of infinite sequence length}

\paragraph{Original Hopfield fitness ($c=-1$).}

The results for the case of three, four and five patterns with
original Hopfield fitness (\ref{original Hopfield fitness}) and
uncorrelated patterns ($X_v=2^{-p}$ for all $v$, corresponding to
random patterns for infinite sequence length) look exactly like those
for two patterns shown in figure \ref{Fig 6.15} (left).

The solutions for the different $x_v$ all coincide. For small mutation
rates $\mu<1/2$, there are again two degenerate solutions for each
$x_v$, which can be combined in multiple ways for the different $x_v$,
yielding any of the patterns or their complementary sequences. At
$\mu=1/2$, there is a second order phase transition, which is a
fitness and degradation threshold.  For small mutation rates, the
population is centred around one of the patterns, say
$\bi{\xi}^{q_c}$, and therefore $y^{q_c}<1/2$, whereas it is
completely random with respect to the other patterns, yielding
$y^q=1/2$ for $q\neq q_c$.

\paragraph{Deviations from the original Hopfield fitness ($c\neq-1$).}

\begin{figure}
\centerline{
\includegraphics[width=\textwidth]
{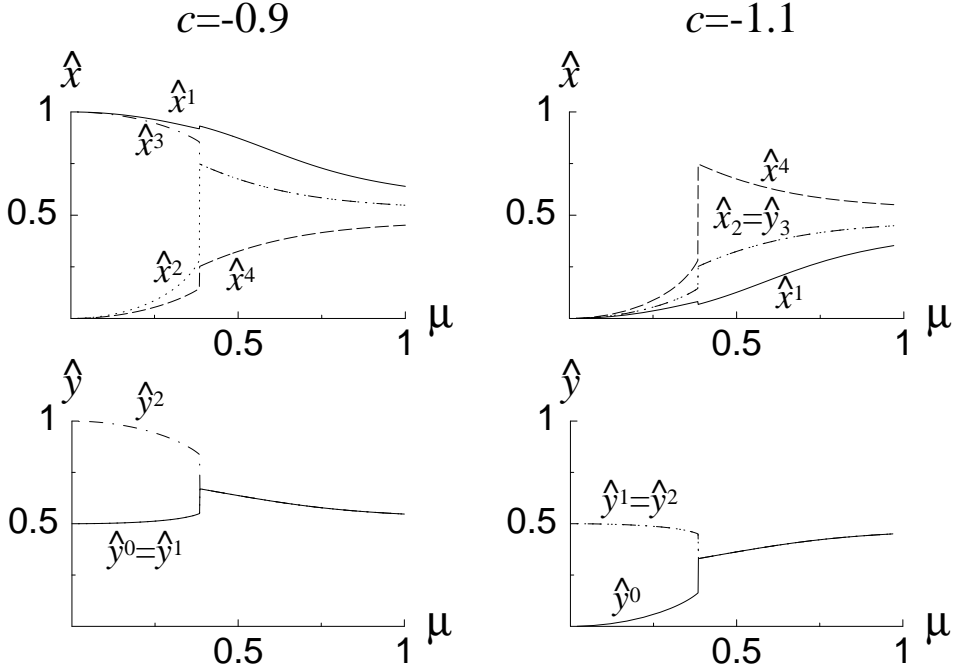}
}
\caption[Ancestral mean partial and specific distances, $\hat{x}_v$ 
         and $\hat{y}^q$, for the quadratic Hopfield-type fitness 
         with three patterns.]
{\label{Fig 6.19}Ancestral mean partial distances $\hat{x}_v$ (top)
  and specific distances $\hat{y}^q$ (bottom) depending on mutation
  rates. The quadratic Hopfield-type fitness (\ref{quadratic symmetric
  Hopfield-type fitness}) with negative epistasis for three patterns
  with $c=-0.9$ (left) and $c=-1.1$ (right) has been used. Results
  correspond to uncorrelated patterns for infinite sequence length
  ($X_v = 1/4 \;\forall v$).}
\end{figure}
Figure \ref{Fig 6.19} shows the ancestral mean partial distances
$\hat{x}_v$ and specific distances $\hat{y}^q$ for the quadratic
Hopfield-type fitness (\ref{quadratic symmetric Hopfield-type
fitness}) with negative epistasis for different values of $c$ in the
case of three patterns. Data for four patterns look very similar (not
shown). As $c$ deviates from $-1$, the solutions for the $x_v$ do not
coincide, and the phase transition becomes a first order fitness
threshold, at which all four partial distances $\hat{x}_v$ jump, but
it is no more a degradation threshold. So contrary to the case of two
patterns, where $\hat{x}_2$ is independent of $c$ and the error
threshold in $\hat{x}_1$ is smoothed out by $c$ deviating from $-1$,
here the threshold concerns all four partial distances $\hat{x}_v$ and
is sharpened to first order by $c\neq-1$.

For $c\neq-1$, the degeneracy between the patterns and their
complements is lifted, and thus for mutation rates below the
threshold, there are only $p+1$ different solutions, correlated with
the patterns for $c<-1$, and with their complements for $c>-1$. For
clarity, only one of the solutions is shown in figure \ref{Fig 6.19}.

\begin{figure}
\centerline{
\includegraphics[width=0.75\textwidth]{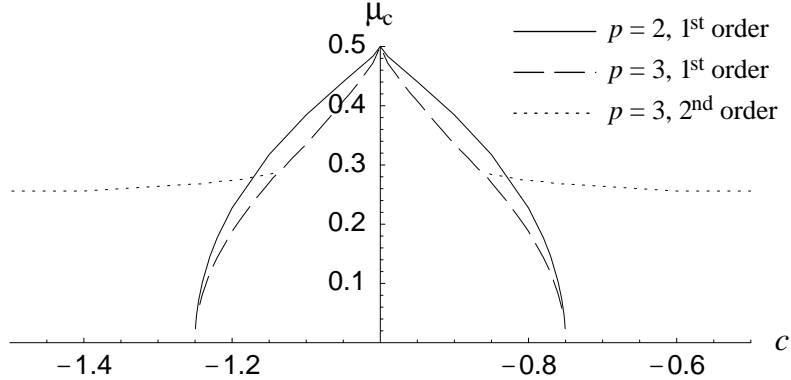}
}
\caption[Dependence of $\mu_c$ on the value of $c$ for the quadratic 
         Hopfield-type fitness with three patterns.]
{\label{Fig 6.20}The critical mutation rate $\mu_c$ depending on the
  value of $c$. The quadratic Hopfield-type fitness (\ref{quadratic 
    symmetric Hopfield-type fitness}) with negative epistasis
  for three and four patterns has been used. Results correspond to
  uncorrelated patterns simulating infinite sequence length.}
\end{figure}

Furthermore, the critical mutation rate decreases with increasing
$|c+1|$. The dependence of the critical mutation rate $\mu_c$ on the
value of $c$ for $p=2,3$ is shown in figure \ref{Fig 6.20}.
At $c=-5/4$ and $c=-3/4$, the critical mutation rate is $\mu_c=0$, and
for values of $c\notin[-5/4,-3/4]$, there are no first order error
thresholds for either $p=2$ or $p=3$. This goes in line with a
different sequence becoming optimal at $\mu=0$ for these values of
$c$.

However, for $p=3$, there is an additional line of second order error
thresholds, that approaches $\mu_c=1/4$ as $|c+1|$ grows. Preliminary
results for $p=4$ indicate, that in that case the second order line
does not occur. It might thus be conjectured that the existence of the
second order error threshold line depends on the number of patterns
being even or odd (remember that for $p=1$ it does exist).

 This is an interesting result, as for all previously
investigated Hopfield-type fitness functions (which are limited to the
original Hopfield fitness and a Hopfield-type truncation selection as
far as the author is aware), the existence of error thresholds has
been reported.

\subsubsection{The case of correlated patterns simulating a finite 
  sequence length}

\begin{figure}
\centerline{
\includegraphics[width=\textwidth]
{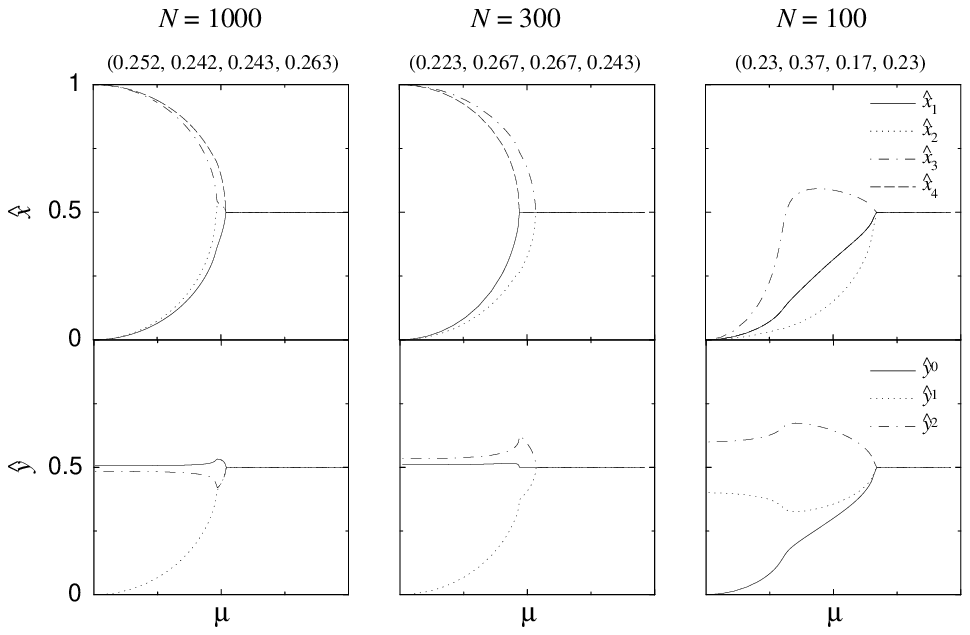}
}
\centerline{
\includegraphics[width=\textwidth]
{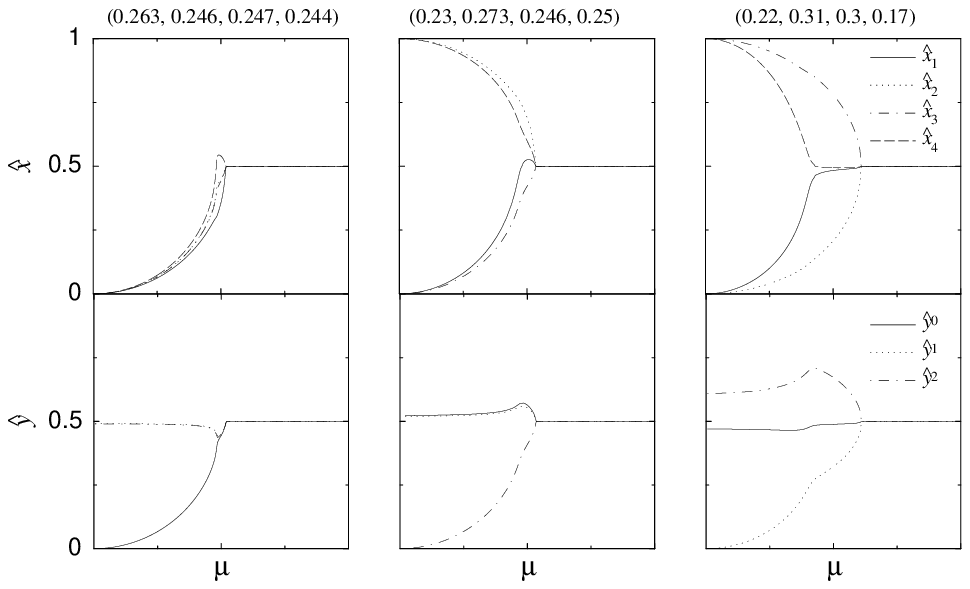}
}
\caption[Ancestral mean partial and specific distances, $\hat{x}_v$ 
         and $\hat{y}^q$, for the original Hopfield fitness with three
         patterns and finite sequence lengths.]
{\label{Fig 6.21}Ancestral mean partial and specific distances,
  $\hat{x}_v$ and $\hat{y}^q$, depending on mutation rates. The
  original Hopfield fitness (\ref{original Hopfield fitness}) for
  three patterns has been used. Results correspond to two typical
  examples of random, but correlated patterns chosen for sequences of
  lengths $N=1000$ (left), $N=300$ (middle) and $N=100$ (right),
  specified at the top of each graph as $(X_1, X_2, X_3, X_4)$.}
\end{figure}
Figure \ref{Fig 6.21} shows some cases of the ancestral mean partial
and specific distances, $\hat{x}_v$ and $\hat{y}^q$, for the original
Hopfield fitness (\ref{original Hopfield fitness}) with three
patterns, which are randomly chosen sequences of finite length. The
correlations between the patterns (and thus the variations of the
$X_v$) are characteristic for the sequence length. The six cases of
patterns shown here are typical examples for the sequence lengths
considered. In the case of long sequences ($N=1000$), the deviations
of the patterns from the infinite sequence limit $X_1=X_2=X_3=X_4$ are
small, and grow with decreasing sequence length. These correlations
that are introduced into the system have the same effect as a choice
of correlated patterns in the case of two patterns, such that the
single critical mutation rate in the case of infinite sequence length
is split up into two critical mutation rates, at each of which two of
the $x_v$ show threshold behaviour. For short sequence length
($N=100$), it can be seen that, particularly at the smaller critical
mutation rate, the threshold is smoothed out.

\begin{figure}
\centerline{
\includegraphics[width=\textwidth]
{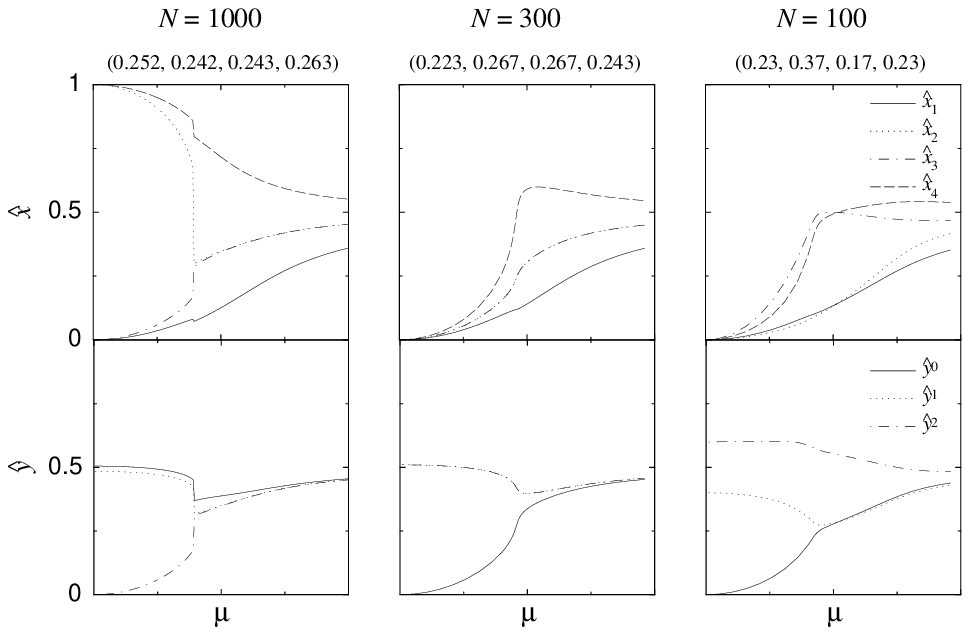}
}
\centerline{
\includegraphics[width=\textwidth]
{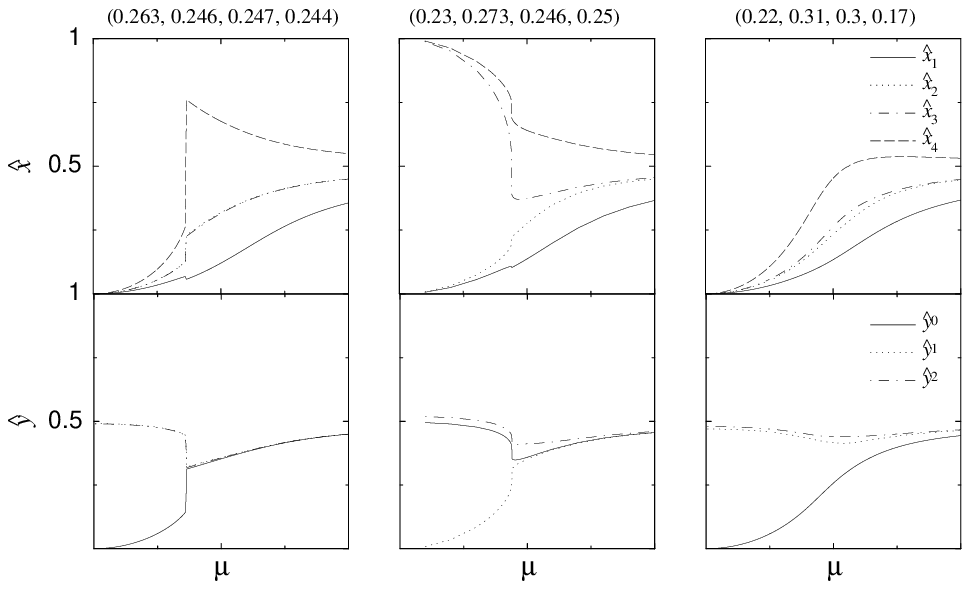}
}
\caption[Ancestral mean partial and specific distances, $\hat{x}_v$ 
         and $\hat{y}^q$, for a quadratic Hopfield-type fitness with 
         three patterns and finite sequence lengths.]
{\label{Fig 6.22}Ancestral mean partial and specific distances,
  $\hat{x}_v$ and $\hat{y}^q$, depending on mutation rates. The
  Hopfield-type fitness (\ref{quadratic symmetric
  Hopfield-type fitness}) with negative epistasis for three patterns
  and $c=-1.1$ has been used. Results correspond to the patterns used
  in figure \ref{Fig 6.21}.}
\end{figure}
In figure \ref{Fig 6.22}, the ancestral mean partial and specific
distances, $\hat{x}_v$ and $\hat{y}^q$, corresponding to the same
patterns as in figure \ref{Fig 6.21} are shown for the quadratic
Hopfield-type fitness (\ref{quadratic symmetric Hopfield-type
fitness}) with negative epistasis and $c=-1.1$. For long sequence
length, these look very similar to the results for infinite sequence
length (cf.\ figure \ref{Fig 6.19}), showing clearly the single first
order phase transition. For shorter sequence lengths, they become more
and more smoothed out, such that at $N=300$, roughly only every other
pattern that was simulated shows an error threshold, whereas for
$N=100$, in the vast majority of cases, there is no threshold. Note
that this effect is present even though the finite sequence length was
only simulated by choosing the patterns accordingly, so it is a
feature of the model with correlated patterns.

\subsection{Summary of the results for Hopfield-type fitness}

In this section, the quadratic symmetric Hopfield-type fitness given
in equation (\ref{quadratic symmetric Hopfield-type fitness}) with
negative epistasis (i.e., positive quadratic term) was investigated
for a small number of patterns. The results for the original Hopfield
fitness ($c=-1$) have been compared with those for the generalised
quadratic Hopfield-type fitness ($c\neq 1$). Furthermore, both
uncorrelated patterns ($X_v=2^{-p}$ for all $v$), corresponding to an
infinite sequence length, and correlated patterns ($X_v\neq2^{-p}$),
simulating a finite sequence length, were considered.  For two
patterns, an analytical treatment was possible, making all values of
the $X_v$ accessible, whereas the case of three or more patterns was
treated numerically due to the larger number of variables. This means
that apart from the uncorrelated choice of pattern ($X_v=2^{-p}$),
which was investigated for three, four and five patterns, only some
correlated combinations for three patterns with $X_v\neq 2^{-p}$ were
investigated, some typical examples of which are shown in section
\ref{The case of three patterns}. The results are summarised as
follows:

\begin{itemize}
\item {\bf Original Hopfield fitness ($c=-1$)}:
\begin{itemize}
\item For {\em uncorrelated patterns}, there is one second order error
  threshold for all $x_v$ at $\mu_c=1/2$ (investigated $p=1,2,3, 4$).
\item For {\em correlated patterns}, there are two second order error
  thresholds, each for half of the $x_v$ (investigated $p=1,2$).
\end{itemize}
\item {\bf Hopfield-type fitness with $c\neq -1$}:
\begin{itemize}
\item For {\em uncorrelated patterns}, there is a first order error
  threshold only on a restricted range of $c$
  ($p=1:$ no first order threshold, $p=2,3: c\in[-5/4, -3/4]$).\\
  For an even number of patterns ($p=1, 3$), there is an additional
  second order error threshold for any $|c+1|\geqslant\Delta c$ ($p=1:
  \Delta c=0$, $p=3: \Delta c \approx 0.145$). This error threshold
  does not exist for the investigated cases of an odd number of
  patterns ($p=2,4$).
\item For {\em correlated patterns}, at $p=1$, there is one second
  order threshold, the other one, which is present for the original
  Hopfield fitness, is smoothed out. At $p=2$, there is up to one
  first order threshold, smoothed out for more strongly correlated
  patterns (corresponding to shorter sequence length).
\end{itemize}
\end{itemize}

The evaluation the Hopfield-type fitness was limited to the cases of
rather small numbers of patterns, simply because an increase in the
number of patterns makes the evaluation more complex. However, the
Hopfield-type fitness was chosen as a potentially realistic fitness
because of its ruggedness that can be tuned by the number of patterns
chosen. The simple cases considered here probably do not show as high
a degree of ruggedness as one would expect for realistic fitness
functions.  However, the results described here already indicate some
features that are common for all numbers of patterns investigated
here, and some that depend on whether the number of patterns is odd or
even. It would be very interesting to establish whether these results
generalise to an arbitrary number of patterns.

Furthermore, the concept of partitioning the set of sites into
subsets, which was introduced to analyse the Hopfield-type fitness, is
very interesting. One could imagine a different interpretation for
this by classifying sites according to the selection strength they
evolve under. Some of the behaviour identified for the Hopfield system
could also occur in such a setting: At intermediate mutation rates,
partially ordered phases could exist, such that sites that evolve
under weak selection have passed their error threshold and the
population is in a phase that is disordered with respect to these
sites, whereas at sites that are subject to strong selection the order
is still maintained.

\section{Conclusion}
\label{Conclusion}

The present work has been concerned with the investigation of a
deterministic mutation--selection model in the sequence space
approach, using a time-continuous formulation.  Important observables
in these mutation--selection models are the population and ancestral
distributions $\bi{p}$ and $\bi{a}$, and means with respect to these
distributions, in particular the population mean fitness $\bar{r}$ and
the ancestral mean genotype $\bi{\hat{x}}$. In equilibrium, $\bi{p}$
is given by the right Perron-Frobenius eigenvector of the
time-evolution operator $\bi{H}$, whereas the ancestral distribution
is given by the product of both right and left PF eigenvectors
$\bi{p}$ and $\bi{z}$, $a_i=z_i p_i$.

Types have been modelled as two-state sequences. As mutation model,
the single step mutation model was used, whereas selection was
modelled by Hopfield-type fitness functions, using the similarity of a
sequence to a number of patterns to determine its fitness. This allows
for a more rugged fitness landscape, and the complexity of the fitness
can be tuned by the number of patterns.

The large number of types that arise in the sequence space approach
have been lumped into classes of types, labelled by the partial
distances $d_v$ introduced in section \ref{Lumping for the
  Hopfield-type fitness} as a generalisation of the Hamming distance.
With this, the maximum principle as developed by \citet{BBBK05} can be
applied to the case of Hopfield-type fitness functions as done in
section \ref{The maximum principle}, see also \citet{Gar05}, treating
two- and four-state sequences.

In section \ref{Error thresholds}, the maximum principle derived in
section \ref{The maximum principle} was used to investigate the
phenomenon of the error threshold. These error thresholds can be
detected with the maximum principle, because the delocalisation of the
population distribution manifests itself as a jump (or at least an
infinite derivative with respect to the mutation rate) of the
ancestral mean genotype $\bi{\hat{x}}$, the maximiser.  Not all
fitness functions give rise to error thresholds, and as the error
thresholds were first described for a model with highly unrealistic
fitness function, it has been argued that they might be an artifact of
this rather than a biologically relevant phenomenon. It is therefore
clearly necessary to investigate more complex fitness functions with
respect to this phenomenon.

Here, quadratic Hopfield-type fitness functions with small numbers of
patterns have been investigated. For the original Hopfield fitness,
the results for all investigated numbers of patterns are identical.
However, if the fitness differs from the original Hopfield fitness,
different behaviours are observed for different numbers of patterns.
In the case of uncorrelated patterns, corresponding to random patterns
chosen for infinite sequence length, the observed features seem to
depend on whether the number of patterns is odd or even. Because for
correlated patterns, only the cases of two and three patterns were
investigated, and found to behave differently, it would be interesting
to see how these results generalise to a higher number of patterns.

In the original Hopfield fitness, error thresholds were observed for
all choices of patterns. This is not true for a generalised
Hopfield-type fitness. For instance, for a Hopfield-type fitness with
positive epistasis no thresholds were observed, going in line with the
results for permutation-invariant fitness. But also for Hopfield-type
fitness functions with negative epistasis, there are not necessarily
any thresholds, if the fitness deviates too much from the original
Hopfield fitness, challenging the commonly held notion that more
complex fitness functions all tend to display error threshold
behaviour. The complexity and ruggedness of the original Hopfield
fitness have been investigated \citep{AGS85a, AGS85b} and found to be
good candidates for realistic fitness functions \citep{Leu87, Tar92}.
However, these results do not necessarily transfer to the generalised
Hopfield-type fitness functions, and therefore it would be very useful
to study these properties of a generalised Hopfield-type fitness
functions to analyse which of these factors are responsible for
generating the thresholds.

\section*{Acknowledgements}

It is my pleasure to thank Uwe Grimm, Michael Baake, Ellen Baake and
Robert Bialowons for helpful discussions and Uwe Grimm for comments on
the manuscript. Support from the British Council and DAAD under the
Academic Research Collaboration Programme, Project no 1213, is
gratefully acknowledged.


\end{document}